\documentclass[11pt]{article}

\usepackage[margin=1in]{geometry}
\usepackage[utf8]{inputenc}
\usepackage[T1]{fontenc}
\usepackage{microtype}
\usepackage{authblk}
\usepackage[numbers,sort&compress]{natbib}
\usepackage{hyperref}
\usepackage{xcolor}
\usepackage{graphicx}
\usepackage{amsmath}
\usepackage{amssymb}
\usepackage{booktabs}
\usepackage{multirow}
\usepackage{xurl}
\usepackage{listings}
\usepackage{enumitem}
\usepackage{subcaption}
\usepackage{caption}

\hypersetup{
  colorlinks=true,
  linkcolor=blue!60!black,
  citecolor=blue!60!black,
  urlcolor=blue!60!black
}

\providecommand{\keywords}[1]{\par\noindent\textbf{Keywords:} #1}

\title{Scientific Code Search at Scale:\\
A Multi-Domain Dataset and Benchmark}

\author[1]{Nishan Pantha}
\author[1]{Pranath Reddy Kumbam}
\author[1]{Sajil Awale}
\author[1]{Pushwitha Krishnappa}
\author[1]{Muthukumaran Ramasubramanian}
\author[1]{Nidhi Jha}
\author[1]{Emily Foshee}
\author[1]{Ankur Kumar}
\author[2]{Rachel Slank}
\author[2]{Ashkbiz Danehkar}
\author[3]{Rahul Ramachandran}

\affil[1]{The University of Alabama in Huntsville (UAH), Huntsville, AL, USA}
\affil[2]{Universities Space Research Association (USRA), USA}
\affil[3]{NASA Marshall Space Flight Center, Huntsville, AL, USA}

\date{}

\begin{document}

\maketitle

\begin{abstract}
Scientists increasingly rely on open-source tools to support their research workflows, yet discovering relevant software among over 600 million GitHub repositories remains challenging. Existing code search benchmarks focus on general software engineering tasks and fail to capture the domain-specific vocabulary and needs of scientific computing. We present a curated \textbf{corpus of 5,264 high-quality, domain-classified scientific repositories} spanning five NASA Science Mission Directorate divisions---Earth Science, Astrophysics, Planetary Science, Heliophysics, and Biological \& Physical Sciences---enriched with cleaned READMEs, extracted topics, and additional context from crawled links. Building on this corpus, we introduce two novel \textbf{information retrieval benchmarks}: (1) a \textbf{repository search benchmark} with 219 expert-curated queries designed by domain scientists, and (2) a large-scale \textbf{code snippet retrieval benchmark} containing 117,950 code snippets and 119,720 queries across seven programming languages. Baseline evaluations on repository search reveal significant performance variation across scientific domains. Code snippet retrieval proves equally challenging, with substantial variation driven by differing documentation practices, coding standards, and programming language conventions across scientific communities. All datasets and benchmarks are publicly released on HuggingFace to support research on scientific tool discovery.
\end{abstract}


% Chapter 1: Introduction, Related Work, and Data Collection

\section{Introduction}
\label{sec:introduction}

The open science movement has transformed how scientific software is developed and shared~\cite{ramachandran2021open}. Researchers increasingly publish code on GitHub, enabling reproducibility and accelerating discovery through reuse~\cite{barker2022introducing, collberg2016repeatability}. However, with over 600 million repositories on GitHub\footnote{\url{https://github.blog/news-insights/octoverse/octoverse-a-new-developer-joins-github-every-second-as-ai-leads-typescript-to-1/}}, finding relevant scientific tools remains challenging. Scientific code discovery differs from general-purpose code search: researchers seek tools for specific data formats (e.g., FITS in astronomy, NetCDF in climate science), particular instruments or missions, and domain-specific algorithms---queries requiring specialized vocabulary that general search systems struggle to capture~\cite{collberg2016repeatability}.

Current platforms like GitHub rely on lexical matching rather than semantic understanding~\cite{husain2019codesearchnet}\footnote{\url{https://github.blog/engineering/architecture-optimization/the-technology-behind-githubs-new-code-search/}}. A researcher searching for \emph{``software to analyze time-series observations from the Kepler mission''} may find nothing, while the same tool indexed under \emph{``exoplanet transit photometry pipeline''} remains undiscovered. This discoverability gap leads to duplicated effort~\cite{hucka2018search}, hinders reproducibility~\cite{garijo2022bestpractices}, and confines effective tools to their originating communities. Existing code search benchmarks---CodeSearchNet~\cite{husain2019codesearchnet}, CoSQA~\cite{huang2021cosqa}, AdvTest~\cite{lu2021codexglue}---draw from general-purpose software, reflect software engineering rather than research queries, and lack scientific vocabulary. Progress requires evaluation infrastructure grounded in real scientific queries and expert relevance judgments.

We present datasets and benchmarks for scientific code discovery. Our contributions include:

\begin{enumerate}[noitemsep]
    \item \textbf{Curated Scientific Repository Corpus}: 5,264 high-quality, domain-classified GitHub repositories spanning five NASA SMD divisions, systematically collected from multiple authoritative sources and enriched with additional context (Appendix~\ref{app:access})

    \item \textbf{Repository Search Benchmark}: 219 expert-curated queries with ground-truth repositories from Earth Science, Astrophysics, and Planetary Science (Appendix~\ref{app:access})

    \item \textbf{Code Snippet Retrieval Benchmark}: A large-scale benchmark with 117,950 code snippets and 119,720 queries across seven programming languages (Python, C, C++, Java, JavaScript, Fortran, and Matlab)

    \item \textbf{Multi-domain Baseline Evaluation}: Evaluation revealing significant performance variation across scientific domains and retrieval approaches (MRR@10 from .18 to .87)

    \item \textbf{Public Release}: All datasets, evaluation scripts, and baselines released on HuggingFace
\end{enumerate}

\section{Related Work}
\label{sec:related}

Code search benchmarks have driven progress in retrieval but target general software engineering. CodeSearchNet~\cite{husain2019codesearchnet} pairs natural language queries with code across six languages; CoSQA~\cite{huang2021cosqa} provides 20,604 query-code pairs from web search; CodeXGLUE~\cite{lu2021codexglue} unifies code understanding tasks; and CoIR~\cite{li2024coir} spans multiple retrieval domains. None address scientific code's specialized terminology, domain-specific data formats, or queries referencing instruments and missions. Scientific software registries---ASCL~\cite{allen2012ascl} for astrophysics, bio.tools~\cite{ison2016biotools} for bioinformatics, SciCrunch~\cite{bandrowski2016scicrunch} for life sciences---catalog software but lack standardized query sets for retrieval evaluation.

On the modeling side, SciBERT~\cite{beltagy2019scibert} and Galactica~\cite{taylor2022galactica} target scientific text, while StarCoder~\cite{li2023starcoder} and CodeLlama~\cite{roziere2023code} provide code foundations, but their training data is predominantly general software. For evaluation, BEIR~\cite{thakur2021beir} and MTEB~\cite{muennighoff2022mteb} offer robust IR benchmarks; we adopt compatible formats to integrate with these pipelines.

\section{Data Collection}
\label{sec:data_collection}

Scientific code is scattered across the web with no centralized index. Unlike general software, which concentrates on platforms like npm or PyPI, scientific tools emerge from research labs, are published alongside papers, and often reside in institutional GitHub organizations unknown to the broader community. Building a representative corpus requires combining multiple discovery channels.

\subsection{Multi-Source Collection Strategy}

We combined four complementary sources (Table~\ref{tab:data_sources}). From \emph{scientific literature}, we mined 142,173 DOIs from the NASA Earth Observations Knowledge Graph~\cite{gerasimov2024bridging}\footnote{\url{https://huggingface.co/datasets/nasa-gesdisc/nasa-eo-knowledge-graph}}, extracting GitHub URLs from 34,712 accessible papers to yield 1,716 repositories. From \emph{institutional organizations}, we enumerated 2,258 public repositories across 53 GitHub organizations identified by domain experts (Appendix~\ref{app:github_orgs}). From \emph{community curation}, we extracted 2,491 GitHub links from the Astrophysics Source Code Library (ASCL)'s 4,398 registry records. From \emph{NASA's Science Discovery Engine}\footnote{\url{https://sciencediscoveryengine.nasa.gov}}, we parsed 546,000 documents to yield 2,553 repositories. After deduplication (removing over 2,100 duplicates across sources), we also developed a keyword-based GitHub API search pipeline\footnote{Available at \url{https://github.com/NASA-IMPACT/github-code-discovery}} for future expansion.

\begin{table}[htbp]
\centering
\begin{tabular}{@{}lrr@{}}
\toprule
\textbf{Source} & \textbf{Repos} & \textbf{\%} \\
\midrule
NASA Science Discovery Engine & 2,199 & 41.77\% \\
Curated GitHub Organizations & 1,384 & 26.3\% \\
Scientific Literature (EO-KG) & 1,339 & 25.44\% \\
ASCL Registry & 320 & 6.07\% \\
\midrule
\textbf{Total} & \textbf{5,242} & 99.58\% \\
\bottomrule
\end{tabular}
\vspace{1ex}
\caption{Data sources for scientific repository corpus. Percentages reflect contribution to the final 5,264 deduplicated repositories after quality filtering.}
\label{tab:data_sources}
\end{table}

\subsection{Quality-Focused Curation Pipeline}

Raw URLs from all sources first underwent standardized processing: we normalized GitHub URLs to base repository format, deduplicated across sources, verified each repository remained active and public, and validated that each contained a non-empty README file. This is critical for our repository search task because information retrieval primarily occurs at the repository documentation level. This initial processing reduced 9,018 raw links to approximately 7,000 unique, accessible repositories with documentation.

To ensure domain relevance, we applied LLM-based classification (described below) to assign each repository to a NASA SMD division, excluding repositories that could not be confidently assigned, including general-purpose tools and software unrelated to NASA science. Finally, we removed repositories with fewer than 50 words in their README, filtering out placeholders and abandoned projects. The resulting corpus contains 5,264 high-quality, domain-relevant repositories.

\subsection{Domain Classification}

We employed an LLM-based classifier using GPT-4.1-mini with few-shot prompting, implemented via pydantic-ai\footnote{\url{https://ai.pydantic.dev/}} for structured output generation. The classifier returns both a division assignment and reasoning for each repository, enabling quality assessment and error analysis. We validated on two 100-sample datasets: 93 Earth Science repositories from EO-KG and 93 Astrophysics repositories from ASCL, both supplemented with 7 common SME-curated negatives. The classifier achieved F1 scores of 0.85 (Earth Science) and 0.93 (Astrophysics); performance was higher on Astrophysics repositories, reflecting the more standardized documentation in that field. The low false negative rate ensures minimal exclusion of genuinely relevant repositories. Appendix~\ref{app:classifier} provides detailed metrics and methodology.

Table~\ref{tab:domain_distribution} shows the resulting distribution. Astrophysics leads (44.05\%), reflecting the field's decades-long culture of open-source development indexed by ASCL. Earth Science follows (39.07\%), driven by the geospatial and climate communities. The remaining domains---Planetary Science (9.91\%), Biological \& Physical Sciences (4.59\%), and Heliophysics (2.35\%)---represent smaller but scientifically important communities whose software remains difficult to discover precisely because of their size. This distribution also reflects biases in our sources: ASCL is a dedicated astrophysics registry and EO-KG draws from Earth Observations literature. We did not incorporate equivalent registries for the underrepresented fields (e.g., HSSI\footnote{\url{https://hssi.hsdcloud.org/}} or HelioPython\footnote{\url{https://heliopython.org/projects/}} for heliophysics); integrating such registries remains future work.

\begin{table}[htbp]
\centering
\begin{tabular}{@{}lrr@{}}
\toprule
\textbf{Domain} & \textbf{Count} & \textbf{\%} \\
\midrule
Astrophysics & 2,319 & 44.05\% \\
Earth Science & 2,057 & 39.07\% \\
Planetary Science & 522 & 9.91\% \\
Biological \& Physical Sciences & 242 & 4.59\% \\
Heliophysics & 124 & 2.35\% \\
\midrule
\textbf{Total} & \textbf{5,264} & 100\% \\
\bottomrule
\end{tabular}
\vspace{1ex}
\caption{Distribution of 5,264 curated repositories across five NASA Science Mission Directorate divisions. Astrophysics and Earth Science dominate due to established open-source cultures and dedicated registries (ASCL, EO-KG).}
\label{tab:domain_distribution}
\end{table}

\subsection{Context Enrichment \& Expansion}

README files alone often lack sufficient semantic signal for effective retrieval. This was particularly evident in domains like Planetary Science, where many repositories contained only installation instructions or generic template documentation rather than descriptions of scientific purpose. Even well-maintained repositories often reference mission-specific instruments without explanation; a repository documenting a ``CRISM data processing pipeline'' may assume familiarity with the Compact Reconnaissance Imaging Spectrometer for Mars, leaving retrieval systems without the context needed to match queries about Mars spectroscopy.

To address these documentation gaps, we enriched each repository through two stages. First, we cleaned each README using an LLM to extract scientifically relevant content and identify key topics, removing boilerplate (installation instructions, licenses, contribution guidelines). Second, we crawled external links from READMEs to gather additional scientific context. Of the 5,264 repositories, 2,943 (56\%) contained at least one high-signal external link (academic publishers, data repositories, documentation platforms), yielding 13,706 links total (4.7 per repository on average). Crawled content was assessed for relevancy using LLM-based scoring and appended to each repository's representation. Implementation details are provided in Appendix~\ref{app:enrichment}.

Table~\ref{tab:dataset_fields} summarizes the fields in the released dataset. We preserve reasoning text for both domain classification and context relevance assessment, so users can audit curation decisions.

\begin{table}[htbp]
\centering
\scriptsize
\begin{tabular}{@{}ll@{}}
\toprule
\textbf{Field} & \textbf{Description} \\
\midrule
\texttt{name} & Repository name \\
\texttt{url} & GitHub repository URL \\
\texttt{description} & Short description from GitHub \\
\texttt{readme} & Raw README content \\
\texttt{readme\_cleaned} & Cleaned README, boilerplate removed \\
\texttt{division} & Predicted NASA SMD division \\
\texttt{division\_reasoning} & LLM reasoning for classification \\
\texttt{topics} & Extracted key topics \\
\texttt{source} & Discovery origin (SDE, Org, EO-KG, ASCL) \\
\texttt{additional\_context} & Content from crawled README links \\
\texttt{..\_reasoning} & Relevance assessment reasoning \\
\bottomrule
\end{tabular}
\vspace{1ex}
\caption{Schema of the released HuggingFace corpus. Each repository record includes raw and cleaned documentation, LLM-assigned domain classification with reasoning, extracted topics, and crawled external context with relevance assessment. The last field is \texttt{additional\_context\_reasoning}.}
\label{tab:dataset_fields}
\end{table}

We use this corpus to construct two retrieval benchmarks: a repository search task (Section~\ref{sec:repo_benchmark}) and a code snippet retrieval task (Section~\ref{sec:code_benchmark}).

% Chapter 2: Repository Search Benchmark

\section{Repository Search Benchmark}
\label{sec:repo_benchmark}

Scientific code discovery typically begins with a research question, not a package name. A climate scientist wondering how to detect deforestation from satellite imagery, or an astronomer seeking tools to analyze data from a specific telescope, expresses their need in the vocabulary of their field---not in the terminology of software documentation. Our first benchmark captures this reality: given a natural language query expressing a scientific computing need, retrieve relevant GitHub repositories from a corpus of scientific software.

This task differs from function-level code retrieval in important ways. Researchers seek entire tools, not code snippets; they evaluate relevance based on README documentation and project descriptions. The retrieval context consists of repository content and descriptions, mirroring the information scientists use when evaluating whether a tool meets their needs.

\subsection{Benchmark Construction}

Unlike benchmarks constructed through automated query generation or crowdsourcing, our repository search benchmark emerged from direct collaboration with practicing scientists. We recruited subject matter experts from Earth Science, Astrophysics, and Planetary Science, asking them to contribute queries representing information needs they or colleagues had actually encountered in their research.

This expert-driven process yielded 219 unique queries. Earth Science dominates with approximately 140 queries, reflecting both the field's diversity---spanning climate modeling, remote sensing, hydrology, and atmospheric science---and the community's active engagement with open-source tools. Astrophysics contributes roughly 50 queries centered on observational data analysis and mission-specific software. Planetary Science adds 30 queries that, while fewer in number, present particular challenges due to mission-specific vocabulary unfamiliar to general retrieval systems. Table~\ref{tab:repo_benchmark_stats} summarizes these statistics.

\begin{table}[htbp]
\centering
\begin{tabular}{@{}lr@{}}
\toprule
\textbf{Statistic} & \textbf{Value} \\
\midrule
Total unique queries & 219 \\
Earth Science queries & $\sim$140 \\
Astrophysics queries & $\sim$50 \\
Planetary Science queries & $\sim$30 \\
Avg. ground truths per query & 2.3 \\
Min ground truths & 1 \\
Max ground truths & 8 \\
\bottomrule
\end{tabular}
\vspace{1ex}
\caption{Repository search benchmark statistics. 219 expert-curated queries span three NASA science domains, with an average of 2.3 ground-truth repositories per query reflecting the multi-relevance nature of scientific tool discovery.}
\label{tab:repo_benchmark_stats}
\end{table}

Annotation followed a \emph{repository-to-query} direction: subject matter experts started from repositories they were already familiar with through their research and formulated queries that those repositories would answer, rather than searching the full corpus for each query. For each query, annotators identified \emph{all} relevant repositories they were aware of, not just a single ``correct'' answer. This design choice reflects scientific reality: multiple tools often address similar needs, and effective retrieval should surface alternatives rather than privileging the most popular option. The resulting benchmark averages 2.3 ground-truth repositories per query, with some queries mapping to as many as eight relevant tools.

The annotation process evolved iteratively. Early rounds assigned single ground truths; subsequent rounds expanded annotations as experts identified additional relevant repositories they had initially overlooked. This iterative refinement improved coverage while maintaining quality through expert judgment rather than automated labeling. We acknowledge the corresponding trade-off: relevant repositories outside an annotator's prior familiarity may be under-represented in the ground truth, a property that favors precision over exhaustive recall.

\subsection{Query Diversity and Multi-Relevance}

The queries in our benchmark span four categories that reflect how scientists actually search for code, illustrated by representative examples in Table~\ref{tab:example_queries}.

\emph{Tool discovery} queries seek specific software packages, often by capability rather than name: ``What Python package can be used to find papers in the Astrophysics Data System?'' These queries test whether retrieval systems can connect capability descriptions to package documentation.

\emph{Workflow questions} frame scientific tasks rather than software needs: ``How do I create map-projected mosaics from CRISM data?'' The researcher knows what they want to accomplish but not which tools enable it---answering such queries requires understanding both the scientific task and available software capabilities.

\emph{Data access} queries focus on finding and retrieving scientific datasets: ``How do I find Earth science datasets across multiple NASA archives?'' These often surface data portal tools and API clients that researchers might not think to search for directly.

\emph{Analysis method} queries seek implementations of scientific techniques: ``How can I track glacier velocity using satellite observations?'' Success requires connecting high-level scientific concepts to specific algorithmic implementations.

\begin{table*}[htbp]
\centering
\begin{tabular}{@{}p{5.5cm}p{2.2cm}p{6cm}@{}}
\toprule
\textbf{Query} & \textbf{Domain} & \textbf{Ground-Truth Repositories} \\
\midrule
Provide Python examples for analysis of Fermi data. & Astrophysics & fermi-lat/AnalysisThreads, ranieremenezes/easyFermi, Fermipy/fermipy, + 5 more \\
\addlinespace
How can I track movement and velocity of glaciers using satellite observations? & Earth Science & nasa-jpl/autoRIFT, nasa-jpl/its\_live, nasa-jpl/its\_live\_production \\
\addlinespace
How do I create PDS labels? & Planetary & NASA-AMMOS/labelocity, NASA-PDS/PLAID, NASA-PDS/mi-label \\
\addlinespace
What pre-trained model can analyze sequences of satellite images to understand changes over time? & Earth Science & NASA-IMPACT/Prithvi-EO-2.0 \\
\bottomrule
\end{tabular}
\vspace{1ex}
\caption{Example queries and ground-truth repositories from the repository search benchmark, illustrating four query categories: tool discovery, workflow questions, data access, and analysis methods. Multi-relevance annotations capture the full ecosystem of tools addressing each scientific need.}
\label{tab:example_queries}
\end{table*}

Consider the Astrophysics query about Fermi gamma-ray data analysis, which maps to eight relevant repositories: the official mission software (fermi-lat/AnalysisThreads), community-developed alternatives (easyFermi, fermipy), and tutorial collections. A retrieval system that returns only the most popular tool fails the researcher who might benefit from alternatives better suited to their specific analysis. This characteristic distinguishes our benchmark from simpler single-answer retrieval tasks and reflects the ecosystem of tools that emerges around major scientific missions and data products. Baseline retrieval results on this benchmark are presented in Section~\ref{sec:repo_results}.

% Chapter 3: Code Snippet Retrieval Benchmark

\section{Code Snippet Retrieval Benchmark}
\label{sec:code_benchmark}

While repository-level search helps scientists discover tools, many research tasks require finding specific implementations: a function that computes spectral density, a class that handles coordinate transformations, or an algorithm for signal detection. Our second benchmark addresses this finer granularity: given a natural language description, retrieve the corresponding code implementation from a corpus of scientific software.\footnote{\url{https://huggingface.co/datasets/nasa-impact/nasa-science-code-benchmark-v0.1}}

We define two complementary retrieval scenarios. In \emph{docstring-to-code} retrieval, the query is a natural language description of functionality---the kind of documentation a scientist might write or search for. In \emph{identifier-to-code} retrieval, the query is a function or class name, testing whether systems can connect abbreviated naming conventions common in scientific code (like \texttt{calc\_snr} for signal-to-noise ratio) to their implementations.

\subsection{Corpus Construction and Composition}

We constructed the code snippet corpus through systematic extraction from our 5,264 scientific repositories, targeting seven programming languages used in scientific computing: Python, C, C++, Java, JavaScript, Fortran, and Matlab.

Using tree-sitter-based parsers from TheVault~\cite{manh2023vault}, we extracted function and class definitions along with their docstrings and identifiers. This process yielded 117,950 unique code snippets and 119,720 queries, summarized in Table~\ref{tab:code_benchmark_stats}.

\begin{table}[htbp]
\centering
\begin{tabular}{@{}lr@{}}
\toprule
\textbf{Statistic} & \textbf{Value} \\
\midrule
Total unique corpus entries & 117,950 \\
Total unique queries & 119,720 \\
Programming languages & 7 \\
Query types & 4 \\
\bottomrule
\end{tabular}
\vspace{1ex}
\caption{Code snippet retrieval benchmark statistics. The corpus comprises 117,950 unique code snippets extracted from scientific repositories using tree-sitter parsers, with 119,720 queries spanning docstring and identifier query types across seven programming languages.}
\label{tab:code_benchmark_stats}
\end{table}

The language distribution in Table~\ref{tab:lang_distribution} reflects the realities of scientific software development. Python dominates at 53.5\%, consistent with its widespread adoption for data analysis and scientific workflows. But the distribution's long tail reveals equally important patterns: C and C++ together account for 26.8\%, representing high-performance simulation code critical to computational science; Java (11.8\%) and JavaScript (4.3\%) capture cross-platform tools and visualization interfaces increasingly common in science.

\begin{table}[htbp]
\centering
\begin{tabular}{@{}lrr@{}}
\toprule
\textbf{Language} & \textbf{Pairs} & \textbf{\%} \\
\midrule
Python & 64,110 & 53.5\% \\
C & 17,149 & 14.3\% \\
C++ & 14,975 & 12.5\% \\
Java & 14,088 & 11.8\% \\
JavaScript & 5,159 & 4.3\% \\
Fortran & 3,586 & 3.0\% \\
Matlab & 653 & 0.5\% \\
\midrule
\textbf{Total} & \textbf{119,720} & 100\% \\
\bottomrule
\end{tabular}
\vspace{1ex}
\caption{Distribution of code retrieval query-snippet pairs across seven programming languages. Python dominates at 53.5\%, while C/C++ (26.8\% combined) represents high-performance scientific computing and Fortran (3\%) captures legacy numerical methods still in active use.}
\label{tab:lang_distribution}
\end{table}

Fortran accounts for 3,586 snippets (3\%), representing numerical methods written decades ago that remain in active use.

The query type distribution in Table~\ref{tab:query_type_distribution} reveals documentation patterns in scientific code. Function-level queries dominate at 78.4\%, reflecting functional programming patterns prevalent in scientific computing. Docstring queries (62.2\%) provide rich semantic content for retrieval, while identifier queries (37.8\%) test mapping of abbreviated domain-specific naming conventions (e.g., \texttt{calc\_snr}, \texttt{get\_wcs}) to implementations.

\begin{table}[htbp]
\centering
\begin{tabular}{@{}lr@{}}
\toprule
\textbf{Query Type} & \textbf{Count} \\
\midrule
Function docstring & 61,083 \\
Function identifier & 32,742 \\
Class docstring & 13,355 \\
Class identifier & 12,540 \\
\midrule
\textbf{Total} & \textbf{119,720} \\
\bottomrule
\end{tabular}
\vspace{1ex}
\caption{Code retrieval pairs by query type. Function-level queries dominate (78.4\%), reflecting functional programming patterns in scientific computing. Docstring queries (62.2\%) provide natural language descriptions, while identifier queries (37.8\%) test retrieval from abbreviated domain-specific names.}
\label{tab:query_type_distribution}
\end{table}

The benchmark follows BEIR~\cite{thakur2021beir} and MTEB~\cite{muennighoff2022mteb} compatible formats, with separate evaluation splits by language and query type. Training and validation splits are available separately. Baseline retrieval results are reported in Section~\ref{sec:code_results}.

% Chapter 4: Baseline Evaluation

\section{Baseline Evaluation}
\label{sec:evaluation}

We evaluate baseline retrieval methods on the repository search benchmark (Section~\ref{sec:repo_benchmark}) and the code snippet retrieval benchmark (Section~\ref{sec:code_benchmark}).

\subsection{Experimental Setup}

Our evaluation compares two fundamental retrieval paradigms: lexical retrieval using BM25 with Okapi BM25 scoring, representing traditional term-matching approaches, and semantic retrieval using dense embedding-based methods that encode queries and documents into continuous vector spaces, enabling similarity matching beyond exact keyword overlap. We evaluate both general-purpose and domain/task-specific embedding models. We assess retrieval quality using standard IR metrics: \textbf{Mean Reciprocal Rank (MRR)} measures how highly the first relevant result ranks; \textbf{Recall@k} measures coverage of relevant results within the top-k; and \textbf{NDCG} evaluates ranking quality accounting for result position. We report metrics at cutoffs @5 and @10.

\subsection{Repository Search Results}
\label{sec:repo_results}

For repository search, we evaluate semantic models at two levels: \texttt{sentence\allowbreak-transformers/\allowbreak all\allowbreak-MiniLM\allowbreak-L6\allowbreak-v2} as a general-purpose embedding baseline and INDUS-Retriever (\texttt{nasa\allowbreak-impact/\allowbreak nasa\allowbreak-smd\allowbreak-ibm\allowbreak-st\allowbreak-v2})~\cite{bhattacharjee-etal-2024-indus} as a domain-specific model trained on a large corpus of scientific literature. We evaluate each method across multiple text representations of repository content: raw README, cleaned README with boilerplate removed, README combined with extracted topics, README with crawled enriched context, and a combined view concatenating all available metadata.

Table~\ref{tab:repo_results} compares retrieval performance across three NASA science domains using lexical (BM25), general semantic (all-MiniLM-L6-v2), and domain-specific (nasa-smd-ibm-st-v2) embedding methods, evaluated over these text representations.

\begin{table*}[htbp]
\centering
\resizebox{\textwidth}{!}{
\begin{tabular}{@{}ll|cccccc|cccccc|cccccc@{}}
\toprule
& & \multicolumn{6}{c|}{\textbf{BM25}} & \multicolumn{6}{c|}{\textbf{all-MiniLM-L6-v2}} & \multicolumn{6}{c}{\textbf{nasa-smd-ibm-st-v2}} \\
& & \multicolumn{2}{c}{MRR} & \multicolumn{2}{c}{Recall} & \multicolumn{2}{c|}{NDCG} & \multicolumn{2}{c}{MRR} & \multicolumn{2}{c}{Recall} & \multicolumn{2}{c|}{NDCG} & \multicolumn{2}{c}{MRR} & \multicolumn{2}{c}{Recall} & \multicolumn{2}{c}{NDCG} \\
\textbf{Text View} & \textbf{Domain} & @5 & @10 & @5 & @10 & @5 & @10 & @5 & @10 & @5 & @10 & @5 & @10 & @5 & @10 & @5 & @10 & @5 & @10 \\
\midrule
\multirow{4}{*}{\texttt{readme}}
& Earth & .220 & .232 & .279 & .370 & .235 & .264 & .343 & .360 & .435 & .552 & .366 & .405 & .417 & .425 & .556 & .621 & .451 & .472 \\
& Astro & .638 & .642 & .709 & .734 & .624 & .631 & .667 & .677 & .662 & .742 & .623 & .648 & .759 & .759 & .832 & .858 & .755 & .762 \\
& Planetary & .123 & .133 & .136 & .210 & .117 & .141 & .173 & .195 & .194 & .391 & .168 & .231 & .203 & .217 & .323 & .442 & .215 & .254 \\
& \textit{Holistic} & .269 & .279 & .324 & .403 & .277 & .302 & .369 & .385 & .437 & .559 & .379 & .418 & .440 & .448 & .567 & .633 & .466 & .487 \\
\midrule
\multirow{4}{*}{\texttt{readme\_cleaned}}
& Earth & .324 & .333 & .434 & .499 & .350 & .371 & .383 & .395 & .527 & .618 & .418 & .447 & .488 & .501 & .634 & .732 & .525 & .556 \\
& Astro & .652 & .662 & .644 & .740 & .587 & .618 & .737 & .747 & .768 & .844 & .711 & .730 & .685 & .690 & .709 & .752 & .651 & .667 \\
& Planetary & .140 & .146 & .198 & .235 & .140 & .153 & .216 & .252 & .247 & .440 & .218 & .286 & .142 & .163 & .188 & .325 & .133 & .181 \\
& \textit{Holistic} & .348 & .357 & .435 & .500 & .358 & .380 & .413 & .428 & .527 & .628 & .436 & .468 & .473 & .486 & .588 & .683 & .493 & .525 \\
\midrule
\multirow{4}{*}{\texttt{readme + topics}}
& Earth & .220 & .232 & .279 & .370 & .235 & .264 & .343 & .360 & .435 & .552 & .366 & .405 & .417 & .425 & .556 & .621 & .451 & .472 \\
& Astro & .638 & .642 & .709 & .734 & .624 & .631 & .667 & .677 & .662 & .742 & .623 & .648 & .759 & .759 & .832 & .858 & .755 & .762 \\
& Planetary & .123 & .133 & .136 & .210 & .117 & .141 & .173 & .195 & .194 & .391 & .168 & .231 & .203 & .217 & .323 & .442 & .215 & .254 \\
& \textit{Holistic} & .269 & .279 & .324 & .403 & .277 & .302 & .369 & .385 & .437 & .559 & .379 & .418 & .440 & .448 & .567 & .633 & .466 & .487 \\
\midrule
\multirow{4}{*}{\shortstack[l]{\texttt{readme +}\\\texttt{additional\_context}}}
& Earth & .210 & .219 & .286 & .351 & .230 & .250 & .337 & .353 & .435 & .552 & .361 & .400 & .420 & .427 & .554 & .602 & .453 & .470 \\
& Astro & .614 & .629 & .604 & .722 & .579 & .614 & .676 & .681 & .694 & .742 & .637 & .650 & .769 & .769 & .821 & .870 & .756 & .772 \\
& Planetary & .168 & .177 & .210 & .284 & .169 & .192 & .173 & .191 & .194 & .354 & .168 & .219 & .128 & .152 & .237 & .405 & .139 & .196 \\
& \textit{Holistic} & .264 & .273 & .323 & .396 & .273 & .296 & .365 & .381 & .442 & .555 & .377 & .414 & .434 & .442 & .553 & .616 & .458 & .479 \\
\midrule
\multirow{4}{*}{\texttt{combined}$^\dagger$}
& Earth & .316 & .326 & .418 & .483 & .342 & .363 & .418 & .424 & .573 & .618 & .454 & .469 & \textbf{.501} & \textbf{.516} & \textbf{.647} & \textbf{.758} & \textbf{.537} & \textbf{.573} \\
& Astro & .748 & .754 & .709 & .755 & .679 & .691 & .815 & .815 & .881 & .892 & .803 & .798 & \textbf{.868} & \textbf{.871} & \textbf{.857} & \textbf{.928} & \textbf{.827} & \textbf{.850} \\
& Planetary & .152 & .161 & .235 & .309 & .159 & .182 & \textbf{.238} & \textbf{.258} & \textbf{.291} & \textbf{.428} & \textbf{.241} & \textbf{.289} & .191 & .218 & .262 & .428 & .190 & .246 \\
& \textit{Holistic} & .359 & .367 & .437 & .500 & .368 & .388 & .453 & .460 & .582 & .634 & .478 & .494 & \textbf{.515} & \textbf{.530} & \textbf{.629} & \textbf{.741} & \textbf{.535} & \textbf{.572} \\
\bottomrule
\end{tabular}

}
\vspace{1ex}
\caption{Repository search benchmark results comparing BM25 (keyword-based), all-MiniLM-L6-v2 (general embeddings), and nasa-smd-ibm-st-v2 (domain-specific embeddings) across text views. \textit{Holistic} aggregates all domains. $^\dagger$\texttt{combined} concatenates \texttt{description}, \texttt{readme\_cleaned}, \texttt{topics}, and \texttt{additional\_context}. Best results per metric in bold. Full results including @1 in Appendix Table~\ref{tab:repo_results_full}.}
\label{tab:repo_results}
\end{table*}

Context enrichment substantially improves retrieval across all methods and domains, suggesting scientific code discovery is fundamentally limited by sparse documentation. INDUS-Retriever achieves the highest overall performance with the \texttt{combined} representation. Astrophysics achieves the best results (MRR@10 .87), benefiting from standardized terminology developed over decades---FITS for data formats, WCS for coordinate systems, photometry for measurement techniques---that appears consistently in both queries and documentation. Earth Science occupies the middle ground (.52 MRR@10), challenged by vocabulary fragmentation: data formats proliferate (NetCDF, GeoTIFF, HDF5, COG, STAC) and a query about ``satellite image analysis'' might match tools documented as ``remote sensing,'' ``Earth observation,'' or ``geospatial processing.'' Planetary Science struggles most (.22 MRR@10), as READMEs tend to assume familiarity with mission-specific instruments (CRISM, HiRISE) and data systems (PDS); the consistent context enrichment gains across all methods in Table~\ref{tab:repo_results} suggest that crawling external documentation is essential for domains with sparse repository-level documentation.

To further probe the benchmark's discriminative power, we evaluated two hybrid retrieval methods that combine lexical and semantic signals: \textbf{Hybrid-RRF}, which fuses BM25 and embedding rankings via Reciprocal Rank Fusion, and \textbf{Hybrid-Rerank}, which applies a cross-encoder to rerank the fused candidates. Table~\ref{tab:hybrid_results} reports results on the \texttt{combined} text view, aggregated across the three benchmarked domains, using \texttt{indus\allowbreak-sde\allowbreak-st\allowbreak-v0.2}~\cite{bhattacharjee-etal-2024-indus} as the embedding model.

\begin{table}[htbp]
\centering
\begin{tabular}{@{}lcccccc@{}}
\toprule
\textbf{Method} & \textbf{MRR@5} & \textbf{MRR@10} & \textbf{R@5} & \textbf{R@10} & \textbf{NDCG@5} & \textbf{NDCG@10} \\
\midrule
BM25              & .359          & .367          & .437          & .500          & .368          & .388 \\
INDUS-SDE (emb.)  & .508          & .518          & .660          & \textbf{.746} & .537          & .566 \\
Hybrid-RRF        & .473          & .487          & .623          & .715          & .502          & .532 \\
Hybrid-Rerank     & \textbf{.522} & \textbf{.528} & \textbf{.668} & .720          & \textbf{.551} & \textbf{.569} \\
\bottomrule
\end{tabular}
\vspace{1ex}
\caption{Hybrid retrieval baselines on the repository search benchmark (\texttt{combined} text view, holistic across Earth Science, Astrophysics, and Planetary Science). Hybrid-RRF fuses BM25 and INDUS-SDE embedding rankings via Reciprocal Rank Fusion; Hybrid-Rerank applies cross-encoder reranking on top of the fused candidates. Best per metric in bold.}
\label{tab:hybrid_results}
\end{table}

Hybrid methods improve substantially over the BM25 lexical baseline (by over 40\% on Recall@10), demonstrating that the benchmark is well-suited for evaluating modern retrieval architectures beyond single-stage lexical or dense methods. Cross-encoder reranking achieves the strongest performance on five of six metrics, modestly improving over the embedding-only baseline. Per-domain Hybrid-Rerank results follow the same domain-difficulty pattern observed earlier: Astrophysics reaches MRR@10 .850, Earth Science .501, and Planetary Science .310, reinforcing that documentation quality across scientific communities drives the largest performance gaps, not retrieval method choice.

\subsection{Code Retrieval Results}
\label{sec:code_results}

While repository search evaluates discovery at the project level, code snippet retrieval tests whether models can locate specific functions and classes within a corpus of 117,950 scientific code snippets spanning seven programming languages. We evaluate five retrieval methods representing distinct paradigms: BM25 as a lexical baseline; two domain-specific semantic models from the INDUS family---\texttt{indus\allowbreak-sde\allowbreak-st\allowbreak-v0.2} and INDUS-Retriever (\texttt{nasa\allowbreak-smd\allowbreak-ibm\allowbreak-st\allowbreak-v2})~\cite{bhattacharjee-etal-2024-indus}; Qwen3-Embedding-0.6B as a general-purpose large language model embedding; and SFR-Embedding-Code-400M\_R as a code-specialized embedding model. Each model encodes docstrings or identifiers as queries against code snippets as documents.

Table~\ref{tab:code_retrieval_results} summarizes retrieval performance at the @10 cutoff across all evaluation dimensions. Full results including @1 and @5 cutoffs are provided in Appendix Table~\ref{tab:consolidated_results}.

\begin{table*}[htbp]
\centering
\resizebox{\textwidth}{!}{
\begin{tabular}{@{}ll|ccc|ccc|ccc|ccc|ccc@{}}
\toprule
& & \multicolumn{3}{c|}{\textbf{BM25}} & \multicolumn{3}{c|}{\textbf{indus-sde-st-v0.2}} & \multicolumn{3}{c|}{\textbf{nasa-smd-ibm-st-v2}} & \multicolumn{3}{c|}{\textbf{Qwen3-Emb.-0.6B}} & \multicolumn{3}{c}{\textbf{SFR-Code-400M\_R}} \\
\textbf{Group} & \textbf{Category} & MRR & Rec. & NDCG & MRR & Rec. & NDCG & MRR & Rec. & NDCG & MRR & Rec. & NDCG & MRR & Rec. & NDCG \\
\midrule
\multirow{5}{*}{\textbf{Division}}
& Astro       & .18 & .26 & .20 & .26 & .41 & .29 & .33 & .49 & .37 & \textbf{.50} & \textbf{.64} & \textbf{.54} & .36 & .48 & .39 \\
& Bio. \& Phys. & .12 & .19 & .14 & .25 & .41 & .29 & .29 & .47 & .33 & \textbf{.57} & \textbf{.71} & \textbf{.61} & .30 & .44 & .34 \\
& Earth       & .22 & .31 & .24 & .31 & .47 & .35 & .39 & .55 & .43 & \textbf{.58} & \textbf{.72} & \textbf{.61} & .43 & .57 & .47 \\
& Helio       & .27 & .36 & .29 & .33 & .49 & .37 & .39 & .54 & .43 & \textbf{.53} & \textbf{.63} & \textbf{.55} & .44 & .56 & .47 \\
& Planetary   & .20 & .30 & .22 & .26 & .41 & .30 & .34 & .52 & .39 & \textbf{.59} & \textbf{.74} & \textbf{.63} & .42 & .58 & .46 \\
\midrule
\multirow{4}{*}{\shortstack[l]{\textbf{Query}\\\textbf{Type}}}
& Class Doc   & .30 & .44 & .34 & .37 & .56 & .42 & .51 & .70 & .55 & \textbf{.74} & \textbf{.88} & \textbf{.78} & .63 & .80 & .67 \\
& Class ID    & .01 & .02 & .01 & .14 & .25 & .16 & .16 & .29 & .19 & \textbf{.25} & \textbf{.41} & \textbf{.29} & .14 & .25 & .16 \\
& Func Doc    & .30 & .43 & .33 & .36 & .54 & .40 & .47 & .66 & .51 & \textbf{.76} & \textbf{.89} & \textbf{.79} & .56 & .72 & .60 \\
& Func ID     & .01 & .01 & .01 & .12 & .22 & .14 & .13 & .24 & .15 & \textbf{.18} & \textbf{.32} & \textbf{.21} & .06 & .12 & .07 \\
\midrule
\multirow{7}{*}{\shortstack[l]{\textbf{Progr.}\\\textbf{Lang.}}}
& C++         & .18 & .28 & .20 & .29 & .47 & .34 & .35 & .54 & .39 & \textbf{.66} & \textbf{.80} & \textbf{.69} & .40 & .55 & .43 \\
& C           & .21 & .30 & .23 & .25 & .41 & .28 & .34 & .52 & .38 & \textbf{.63} & \textbf{.79} & \textbf{.67} & .39 & .54 & .42 \\
& Fortran     & .18 & .27 & .20 & .28 & .45 & .32 & .30 & .47 & .34 & \textbf{.48} & \textbf{.60} & \textbf{.51} & .34 & .45 & .36 \\
& Java        & .05 & .07 & .05 & .20 & .35 & .23 & .22 & .37 & .26 & \textbf{.33} & \textbf{.49} & \textbf{.37} & .18 & .30 & .21 \\
& Javascript  & .27 & .37 & .30 & .30 & .45 & .33 & .36 & .51 & .40 & \textbf{.62} & \textbf{.74} & \textbf{.65} & .45 & .54 & .47 \\
& Matlab      & .04 & .04 & .04 & .15 & .22 & .17 & .21 & .32 & .24 & \textbf{.26} & \textbf{.35} & \textbf{.28} & .21 & .28 & .23 \\
& Python      & .21 & .30 & .23 & .28 & .43 & .32 & .37 & .53 & .41 & \textbf{.53} & \textbf{.66} & \textbf{.56} & .41 & .54 & .44 \\
\midrule
\textbf{Holistic} & Overall & .19 & .27 & .21 & .27 & .43 & .31 & .34 & .51 & .38 & \textbf{.54} & \textbf{.68} & \textbf{.58} & .38 & .51 & .41 \\
\bottomrule
\end{tabular}
}
\vspace{1ex}
\caption{Code snippet retrieval results at @10 cutoff comparing five methods---BM25 (lexical), two INDUS domain-specific models, Qwen3-Embedding-0.6B (general-purpose LLM), and SFR-Embedding-Code-400M\_R (code-specialized)---across scientific divisions, query types, and programming languages. Best results per row in bold. Full results with @1 and @5 cutoffs in Appendix Table~\ref{tab:consolidated_results}.}
\label{tab:code_retrieval_results}
\end{table*}

Qwen3-Embedding-0.6B achieves the strongest overall performance (MRR@10 .54, Recall@10 .68, NDCG@10 .58)---nearly double BM25 and substantially outperforming domain-specific INDUS models. Qwen3-Embedding-0.6B, a competitive model on the MTEB leaderboard~\cite{muennighoff2022mteb} built on a decoder-based language model with multi-stage contrastive training, shows strong performance here suggesting that rich general text understanding transfers effectively to scientific code retrieval even without domain-specific fine-tuning. This contrasts with repository search where INDUS-Retriever led, indicating that scientific-text pretraining benefits document-level retrieval but does not transfer as directly to code-level tasks. SFR-Embedding-Code-400M\_R ranks second, reinforcing the value of code-aware representations.

The largest gap appears across query types. Docstring-based queries achieve strong retrieval performance, with Qwen3 reaching MRR@10 of .76 for function docstrings and .74 for class docstrings. Identifier-based queries, however, prove far more challenging: even the best model achieves only .25 MRR@10 for class identifiers and .18 for function identifiers. BM25 essentially fails on identifier queries (MRR@10 of .01), as function and class names in scientific code often use abbreviated, domain-specific naming conventions (e.g., \texttt{calc\_wcs\_transform}, \texttt{pds\_reader}) that share minimal lexical overlap with natural-language descriptions. This gap highlights a fundamental challenge: while scientists can describe what code does in natural language, retrieving code by its programmatic identifiers requires understanding naming conventions that vary across communities and lack standardization.

Division-level results reverse the pattern from repository search. Planetary Science now leads (MRR@10 .59), while Astrophysics trails (.50)---the opposite of repository search. This reflects differing documentation cultures: Astrophysics benefits from excellent project-level documentation driven by ASCL, but inline docstring coverage varies across its large corpus (44\% of repositories). Planetary Science, despite sparse READMEs, enforces strict PDS software engineering standards including mandatory code reviews and API documentation generation, prioritizing function-level documentation. Its smaller, more curated corpus (9.9\% of repositories) also concentrates higher-quality snippets, likely contributing to stronger retrieval performance. Earth Science occupies a middle ground, where metadata-centric conventions (CF standards, STAC specifications) encourage structured function-level documentation. Programming language variation is also substantial, ranging from C++ (.66) and C (.63) at the top to Java (.33) and Matlab (.26) at the bottom. This ordering contrasts with general code retrieval benchmarks where Python typically achieves the highest retrieval scores, suggesting that the variation here is driven by corpus-specific factors---the documentation quality and coding conventions of the particular scientific repositories in each language---rather than inherent language properties. Python (.53) falls in the middle despite being the most represented language, likely because its large and heterogeneous corpus spans both well-documented libraries and minimal analysis scripts.

\subsection{Implications}

These results carry clear implications for scientific code search research. First, documentation culture matters at both repository and code levels, but README quality and inline docstring quality are distinct dimensions that do not necessarily correlate---domain adaptation must account for both. Second, context enrichment provides measurable gains for repository search across all methods, suggesting that documentation augmentation is a viable improvement strategy independent of the retrieval model. Third, code-aware representations are essential for code retrieval: domain-specific scientific-text embeddings do not transfer effectively to code-level search, where general-purpose language model embeddings substantially outperform them. Fourth, the stark docstring-vs-identifier gap (.76 vs.\ .25 MRR@10) and cross-domain performance variations make these benchmarks well-suited for evaluating agentic retrieval systems capable of iterative query refinement, context gathering, and multi-step reasoning---approaches that may substantially outperform the single-pass methods evaluated here. The benchmarks are modular and can also serve as the retrieval component in retrieval-augmented generation (RAG) pipelines, with downstream generation quality evaluated separately as a complementary study.

% Chapter Final: Limitations, Future Work, Ethics, and Conclusion

\subsection{Limitations}
\label{sec:limitations}

We acknowledge several limitations: (1)~\textbf{Domain coverage imbalance}: Heliophysics (124 repos) and Biological \& Physical Sciences (242 repos) are underrepresented, reflecting both the current state of open-source scientific software and gaps in our collection methodology; models evaluated on larger domains may not generalize to these smaller communities. (2)~\textbf{English-only queries}: all queries and documentation are in English, limiting applicability to multilingual scientific communities where researchers may describe their needs in other languages or use non-English technical vocabulary. (3)~\textbf{Temporal dynamics}: repositories evolve continuously---new versions are released, documentation is updated, and projects are archived---meaning the benchmark captures a snapshot that may require periodic updates to remain representative. (4)~\textbf{Expert availability}: SME-curated queries cover three of five domains (Earth Science, Astrophysics, Planetary Science), with limited coverage of Heliophysics and Biological \& Physical Sciences; expanding expert participation to these domains would strengthen the benchmark's cross-domain evaluation capabilities.

\subsection{Future Work}

Our benchmarks open several research directions. First, expanding the dataset to better cover Heliophysics and Biological \& Physical Sciences---currently underrepresented with only 124 and 242 repositories respectively---by incorporating domain-specific registries such as HSSI and HelioPython for heliophysics, and bio.tools for biological sciences, would enable more balanced evaluation across all five NASA SMD divisions. Second, agentic retrieval workflows are a natural next step: systems that iteratively refine queries, gather additional context from external sources, and reason over multi-hop evidence could substantially outperform the single-pass methods evaluated here. The difficulty of identifier-based retrieval (.25 MRR@10 vs.\ .76 for docstrings) and cross-domain performance gaps make these benchmarks particularly well-suited for evaluating such multi-step retrieval agents. A closely related direction is end-to-end evaluation of retrieval-augmented generation (RAG) pipelines built atop these benchmarks, where downstream task quality is measured alongside retrieval metrics; we plan this generation evaluation as a follow-up study. Third, domain-adaptive fine-tuning of embedding models on scientific vocabulary could improve performance, and investigating cross-domain transfer---how models trained on well-represented domains generalize to smaller scientific communities---remains an open question.

\subsection{Ethical Considerations}
\label{sec:ethics}

All data in our corpus is sourced from publicly available GitHub repositories and open scientific registries; no private or personal data was collected, and no human subjects were involved. We respect the licensing terms of individual repositories: the publicly released repository corpus redacts all original repository content---raw and cleaned READMEs, extracted topics, and crawled additional context---and contains only our metadata, annotations, and provenance fields under the CC-BY-4.0 license. We provide replication scripts so that users can regenerate the redacted fields by fetching content directly from source repositories, where the original repository licenses apply. We acknowledge domain distribution bias: Astrophysics and Earth Science are over-represented due to the maturity of their open-source ecosystems and the availability of domain-specific registries (ASCL, EO-KG). We disclose the use of LLMs (GPT-4.1-mini) for domain classification and README cleaning; all prompts, validation methodology, and classifier performance metrics are provided in the Appendix to enable reproducibility and scrutiny of these automated decisions. The benchmark's scientific focus limits direct misuse potential, though models evaluated on scientific text may exhibit inflated performance if pre-trained on overlapping corpora. We encourage researchers to report training data overlap when using these benchmarks.

\section{Conclusion}
\label{sec:conclusion}

We introduced the first benchmarks specifically designed for scientific code discovery, addressing a critical gap between existing code search benchmarks and the needs of the scientific community. Our contributions comprise a curated corpus of 5,264 domain-classified scientific repositories spanning five NASA SMD divisions, a repository search benchmark with 219 expert-curated queries grounded in real scientific information needs, and a code snippet retrieval benchmark with 117,950 code snippets and 119,720 queries across seven programming languages. Baseline evaluations reveal that repository search accuracy varies significantly across scientific domains, with context enrichment providing consistent gains that underscore the importance of documentation augmentation. Code snippet retrieval shows that LLM-based embeddings (Qwen3-Embedding-0.6B, .54 MRR@10)---a competitive model on the MTEB leaderboard built on decoder-based language model foundations with multi-stage contrastive training---substantially outperform both smaller domain-specific and code-specialized models, suggesting that strong general text understanding transfers effectively to scientific code retrieval. Retrieval performance drops sharply when code lacks descriptive documentation and relies on abbreviated, non-informative identifiers (.25 vs.\ .76 MRR@10), exposing how prevalent terse naming conventions in scientific codebases remain a major barrier to discoverability. These findings highlight that improving scientific code discovery requires advances on multiple fronts: richer documentation practices, domain-aware retrieval models, and benchmarks that capture the distinctive challenges of scientific software. All datasets, evaluation scripts, and baselines are publicly released on HuggingFace to enable reproducible research on scientific code discovery.

\section*{Acknowledgments}

This work was supported by NASA Grant 80MSFC22M004.

\bibliographystyle{plainnat}
\bibliography{main}

% Appendix

\clearpage
\appendix

\section{Query Examples}
\label{app:queries}

Table~\ref{tab:appendix_queries} provides additional example queries from each scientific domain in the repository search benchmark.

\begin{table*}[htbp]
\centering
\begin{tabular}{@{}p{6cm}p{2.2cm}p{5.2cm}@{}}
\toprule
\textbf{Query} & \textbf{Domain} & \textbf{Ground-Truth Repositories} \\
\midrule
What toolkit can be used to build a single mosaic FITS file image from multiple FITS image files? & Astrophysics & Caltech-IPAC/Montage \\
\addlinespace
Provide a Python tool for checking transit signals in Kepler monitoring data. & Astrophysics & SSDataLab/vetting, nasa/K2CE, lightkurve/lightkurve \\
\addlinespace
How do I orchestrate large-scale processing pipelines for Earth observation datasets? & Earth Science & nasa/cumulus, nasa/cumulus-dashboard \\
\addlinespace
What service can validate the quality and completeness of metadata descriptions for my satellite datasets? & Earth Science & NASA-IMPACT/QuARC \\
\addlinespace
How do I render an image taken by Curiosity and overlay its viewing geometry in a 3D scene? & Planetary & NASA-AMMOS/CameraModelUtilsJS \\
\addlinespace
How do I bundle SPICE kernels of Europa Clipper's trajectory during a flyby of Europa? & Planetary & NASA-PDS/naif-pds4-bundler \\
\bottomrule
\end{tabular}
\vspace{1ex}
\caption{Additional example queries from the repository search benchmark.}
\label{tab:appendix_queries}
\end{table*}

\section{GitHub Organization List}
\label{app:github_orgs}

Table~\ref{tab:github_orgs} lists the 53 GitHub organizations included in our curated organization crawl: 24 Earth Science, 12 Planetary Science, 11 Astrophysics, 3 Heliophysics, and 3 spanning multiple divisions.

\begin{table*}[htbp]
\centering
\scriptsize
\begin{tabular}{@{}p{7.5cm}p{5.5cm}l@{}}
\toprule
\textbf{Organization Name} & \textbf{GitHub URL} & \textbf{Division} \\
\midrule
Advanced Rapid Imaging and Analysis (ARIA) & github.com/aria-jpl & Earth Science \\
Alaska Satellite Facility (ASF) & github.com/asfadmin & Earth Science \\
Caltech IPAC Software & github.com/Caltech-IPAC & Astrophysics \\
Integrated Software for Imagers and Spectrometers v3 (ISIS3) & github.com/DOI-USGS/ISIS3 & Planetary Science \\
Earth Observing System Data and Information System (EOSDIS) & github.com/eosdis-nasa & Earth Science \\
Fermi Gamma-Ray Space Telescope & github.com/fermi-lat & Astrophysics \\
GHRC DAAC & github.com/ghrcdaac & Earth Science \\
HDF-EOS Tools and Information Center & github.com/hdfeos & Earth Science \\
HPDE & github.com/hpde & Heliophysics \\
Machine Learning and Instrument Autonomy & github.com/JPLMLIA & Multiple \\
JPL Ephemeris Reader for Go & github.com/mshafiee/jpl & Planetary Science \\
NASA & github.com/nasa & Multiple \\
NASA Advanced Multi-Mission Operations System (AMMOS) & github.com/NASA-AMMOS & Planetary Science \\
NASA Common Metadata Repository (CMR) & github.com/nasa-cmr & Earth Science \\
NASA Cryospheric Sciences Laboratory & github.com/NASA-Cryospheric-Sciences-Laboratory & Earth Science \\
NASA DEVELOP & github.com/NASA-DEVELOP & Earth Science \\
NASA Fornax & github.com/nasa-fornax & Astrophysics \\
General Coordinates Network (GCN) & github.com/nasa-gcn & Astrophysics \\
NASA GIBS & github.com/nasa-gibs & Earth Science \\
NASA IMPACT & github.com/NASA-IMPACT & Earth Science \\
NASA Jet Propulsion Laboratory (JPL) & github.com/nasa-jpl & Planetary Science \\
NASA Lambda & github.com/nasa-lambda & Astrophysics \\
NASA Land Information System (LIS) & github.com/NASA-LIS & Earth Science \\
NASA NAVO & github.com/NASA-NAVO & Astrophysics \\
NASA Center for Climate Simulation (NCCS) & github.com/nasa-nccs-cds & Earth Science \\
NASA GSFC Data Science Group & github.com/nasa-nccs-hpda & Earth Science \\
NASA Openscapes & github.com/nasa-openscapes & Earth Science \\
NASA PDS & github.com/nasa-pds & Planetary Science \\
PDS Engineering Node & github.com/nasa-pds-engineering-node & Planetary Science \\
NASA Planetary Science & github.com/NASA-Planetary-Science & Planetary Science \\
NASA Student Airborne Research Program (SARP) & github.com/NASA-SARP & Earth Science \\
Salinity and Stratification at the Sea Ice Edge (SASSIE) & github.com/NASA-SASSIE & Earth Science \\
NASA ARSET & github.com/NASAARSET & Earth Science \\
NASA Datanauts & github.com/NASADatanauts & Multiple \\
Planetary Spectrum Generator (PSG) & github.com/nasapsg & Planetary Science \\
NASA Psyche & github.com/NASAPsyche & Planetary Science \\
NASA WorldWind & github.com/NASAWorldWind & Earth Science \\
NASA Ames Stereo Pipeline & github.com/\allowbreak NeoGeographyToolkit/\allowbreak StereoPipeline & Planetary Science \\
National Snow and Ice Data Center (NSIDC) & github.com/nsidc & Earth Science \\
ANISE (Attitude, Navigation, Instrument, Spacecraft, Ephemeris) & github.com/nyx-space/anise & Planetary Science \\
ORNL DAAC & github.com/ornldaac & Earth Science \\
NASA Ames PAH IR Spectroscopic Database (PAHdb) & github.com/PAHdb & Astrophysics \\
PO.DAAC & github.com/podaac & Earth Science \\
PUNCH Mission & github.com/punch-mission & Heliophysics \\
SEADAS & github.com/seadas & Earth Science \\
SERVIR Amazonia & github.com/SERVIR-Amazonia & Earth Science \\
SnowEx & github.com/SnowEx & Earth Science \\
Space Telescope Science Institute & github.com/spacetelescope & Astrophysics \\
Space Science Data Lab & github.com/SSDataLab & Astrophysics \\
Space Weather Science Operations Center (SWxSOC) & github.com/swxsoc & Heliophysics \\
Gamma-ray Data Tools & github.com/USRA-STI & Astrophysics \\
Mars Target Encyclopedia (MTE) & github.com/wkiri/MTE & Planetary Science \\
EXOplanet Transit Interpretation Code (EXOTIC) & github.com/rzellem/EXOTIC & Astrophysics \\
\bottomrule
\end{tabular}
\vspace{1ex}
\caption{Complete list of 53 curated GitHub organizations included in the corpus, organized by NASA Science Mission Directorate division.}
\label{tab:github_orgs}
\end{table*}

\section{Domain Classification}
\label{app:classifier}

We explored both classical machine learning and LLM-based approaches for domain classification before selecting the production classifier.

\textbf{Baseline Classifier.} As a baseline, we generated sentence embeddings using INDUS-Retriever~\cite{bhattacharjee-etal-2024-indus} and applied a K-Nearest Neighbors classifier with cosine distance as the similarity metric. We evaluated on two datasets constructed from our source data:

\begin{itemize}[noitemsep]
    \item \textbf{Dataset-1}: 2,258 positive samples from the SME-curated organization list, plus 207 negative samples (7 SME-curated and 200 synthetically generated using Gemma3-4B)
    \item \textbf{Dataset-2}: 1,716 positive samples from the EO Knowledge Graph DOI links, plus 207 negative samples
\end{itemize}

The baseline achieved high metrics (F1 $>$ 0.98 on both datasets), but showed signs of overfitting and failed to generalize when applied to repositories discovered through keyword-based search. The limited negative samples and domain-specific embedding space contributed to this overfitting behavior.

\textbf{LLM Classifier.} The classifier uses pydantic-ai for few-shot LLM-based classification with structured output generation. The agent is configured to return one of five NASA SMD divisions if the repository is deemed relevant to NASA science, or ``Not a NASA Division'' otherwise, along with reasoning for the classification decision.

We selected OpenAI models for their accessible API and reliable structured output generation. We evaluated multiple models (gpt-4o-mini, gpt-o4-mini, gpt-o3, gpt-4.1, gpt-4.1-mini) on two evaluation datasets of 100 samples each: one comprising 93 Earth Science repositories from EO-KG, and another with 93 Astrophysics repositories from ASCL. Both datasets include 7 common SME-curated negative samples. Performance was notably higher on Astrophysics repositories (from ASCL) compared to Earth Science (from EO-KG), reflecting the more standardized terminology in astrophysics documentation. Initial experiments showed low recall due to overly conservative classification; we improved results by adjusting the prompt to reduce false negatives. The final classifier using gpt-4.1-mini achieved F1 scores of 0.93 on Astrophysics and 0.85 on Earth Science samples with low false negative rate, ensuring minimal exclusion of genuinely relevant repositories (Tables~\ref{tab:llm_eo} and~\ref{tab:llm_ascl}).

The classification prompt includes task-specific instructions, evaluation guidelines aligned with the NASA Science Discovery Engine document, and few-shot examples for Earth Science and Astrophysics divisions. The full prompt with detailed evaluation criteria and few-shot examples is provided in Appendix~\ref{app:division_prompt}.

Table~\ref{tab:llm_eo} and Table~\ref{tab:llm_ascl} present detailed performance metrics across multiple OpenAI models on our two evaluation datasets.

\begin{table}[htbp]
\centering
\begin{tabular}{@{}lrrrrr@{}}
\toprule
\textbf{Model} & \textbf{Precision} & \textbf{Recall} & \textbf{Accuracy} & \textbf{F1} & \textbf{FNR} \\
\midrule
gpt-4o-mini & 0.956 & 0.699 & 0.690 & 0.808 & 0.301 \\
gpt-o4-mini & 0.970 & 0.688 & 0.690 & 0.805 & 0.312 \\
gpt-o3 & 0.986 & 0.742 & 0.750 & 0.847 & 0.258 \\
gpt-4.1 & 0.947 & 0.763 & 0.740 & 0.845 & 0.237 \\
gpt-4.1-mini & 0.936 & 0.785 & 0.750 & \textbf{0.854} & 0.215 \\
\bottomrule
\end{tabular}
\vspace{1ex}
\caption{LLM classifier performance on EO-KG Earth Science repositories (n=100).}
\label{tab:llm_eo}
\end{table}

\begin{table}[htbp]
\centering
\begin{tabular}{@{}lrrrrr@{}}
\toprule
\textbf{Model} & \textbf{Precision} & \textbf{Recall} & \textbf{Accuracy} & \textbf{F1} & \textbf{FNR} \\
\midrule
gpt-4o-mini & 0.963 & 0.850 & 0.830 & 0.903 & 0.151 \\
gpt-o4-mini & 0.988 & 0.914 & 0.910 & \textbf{0.950} & 0.086 \\
gpt-o3 & 0.977 & 0.903 & 0.890 & 0.939 & 0.097 \\
gpt-4.1 & 0.966 & 0.903 & 0.880 & 0.933 & 0.097 \\
gpt-4.1-mini & 0.955 & 0.903 & 0.870 & 0.928 & 0.097 \\
\bottomrule
\end{tabular}
\vspace{1ex}
\caption{LLM classifier performance on ASCL Astrophysics repositories (n=100).}
\label{tab:llm_ascl}
\end{table}

\section{Context Enrichment Details}
\label{app:enrichment}

README files alone often lack sufficient semantic signal for effective retrieval. This was particularly evident in domains like Planetary Science, where many repositories contained only installation instructions or generic template documentation rather than descriptions of scientific purpose. Even well-maintained repositories often reference mission-specific instruments without explanation; a repository documenting a ``CRISM data processing pipeline'' may assume familiarity with the Compact Reconnaissance Imaging Spectrometer for Mars, leaving retrieval systems without the context needed to match queries about Mars spectroscopy.

To address these documentation gaps, we enriched the contextual representation of each repository through a two-stage process. First, we cleaned each README using an LLM to extract scientifically relevant content and identify key topics, removing boilerplate text such as installation instructions, license information, and contribution guidelines. This produces a \texttt{readme\_cleaned} field that focuses on the repository's scientific purpose, along with extracted \texttt{topics} that capture the primary themes and applications.

Second, we expanded contextual representation by crawling external links present in README files. Using crawl4ai\footnote{\url{https://github.com/unclecode/crawl4ai}} for bulk web crawling, we first filtered out non-informative links such as social media profiles, contact pages, about-us sections, and generic image URLs. For each remaining link, we retrieved content and used LLM-based structured generation to assess whether the content provided information relevant to the repository's scientific purpose. Relevant information was extracted and stored as \texttt{additional\_\allowbreak{}context}, with the extraction reasoning preserved in \texttt{additional\_\allowbreak{}context\_\allowbreak{}reasoning} to enable quality assessment.

From the 5,264 repositories, we identified 2,943 (56\%) containing at least one high-signal external link---those pointing to academic publishers (arxiv.org, doi.org, ieee.org), data repositories (zenodo.org, figshare.com, data.nasa.gov), or documentation platforms (readthedocs.io, huggingface.co). Non-informative links such as social media profiles, CI/CD badges, and recursive repository links were filtered out. In total, 13,706 high-signal links were extracted, averaging 4.7 per repository.

For multi-format content extraction (PDF, HTML, DOCX), we used Docling~\cite{Docling} to convert all content to normalized markdown. Crawled content was assessed for relevancy using a scoring agent evaluating six dimensions: topic alignment, content depth, evidence quality, methodological relevance, recency, and scope. Content scoring above a 0.5 threshold was appended to the repository's enriched representation.

\section{Annotation Guidelines}
\label{app:annotation}

Subject matter experts were provided with the following guidelines for creating queries:

\begin{enumerate}
    \item \textbf{Authenticity}: Write queries representing real information needs you or colleagues have encountered in your research.

    \item \textbf{Answerability}: Ensure the query can be answered by examining repository README files and descriptions.

    \item \textbf{Specificity}: Include domain-specific terminology where appropriate (e.g., instrument names, data formats, mission names).

    \item \textbf{Completeness}: Identify ALL relevant repositories from the corpus, not just the most obvious one.

    \item \textbf{Diversity}: Cover different query types:
    \begin{itemize}
        \item Tool discovery (``What package does X?'')
        \item Workflow questions (``How do I accomplish Y?'')
        \item Data access (``How do I get data from Z?'')
        \item Analysis methods (``What tools implement algorithm W?'')
    \end{itemize}
\end{enumerate}

\section{Code Retrieval Benchmark Datasets}
\label{app:code_benchmark}

We constructed four code retrieval benchmark datasets from public NASA repositories to evaluate semantic code search capabilities. All parsing was performed using tree-sitter-based parsers from TheVault codebase~\cite{manh2023vault}, providing consistent extraction across seven programming languages: Python, C, C++, Java, JavaScript, Fortran, and Matlab.

The four dataset subsets serve complementary evaluation purposes:
\begin{itemize}[noitemsep]
    \item \textbf{Function-Code-Docstring}: Retrieval using function docstrings as queries
    \item \textbf{Class-Code-Docstring}: Retrieval using class docstrings as queries
    \item \textbf{Function-Code-Identifier}: Retrieval using function names as queries (with masked identifiers)
    \item \textbf{Class-Code-Identifier}: Retrieval using class names as queries (with masked identifiers)
\end{itemize}

The docstring-based datasets evaluate semantic code search where natural language descriptions serve as queries. The identifier-based datasets use function and class names as queries while masking these identifiers in the code body, preventing trivial keyword matching and forcing retrieval systems to rely on semantic understanding of code structure.

\subsection{Function-Code-Docstring Dataset}

We extracted functions with associated docstrings from Python source code across public NASA repositories. The initial scan identified over 2.5 million functions across seven languages. Table~\ref{tab:func_docstring_stats} presents the extraction statistics.

\begin{table}[htbp]
\centering
\begin{tabular}{@{}lrrr@{}}
\toprule
\textbf{Language} & \textbf{Total Functions} & \textbf{With Docstrings} & \textbf{Percentage} \\
\midrule
C & 407,637 & 178,904 & 43.89\% \\
Java & 223,585 & 92,534 & 41.39\% \\
Python & 823,958 & 424,497 & 51.52\% \\
Fortran & 143,463 & 80,638 & 56.21\% \\
JavaScript & 173,228 & 51,767 & 29.88\% \\
Matlab & 36,404 & 5,897 & 16.20\% \\
C++ & 771,189 & 206,404 & 26.76\% \\
\midrule
\textbf{Total} & \textbf{2,579,464} & \textbf{1,040,641} & \textbf{40.34\%} \\
\bottomrule
\end{tabular}
\vspace{1ex}
\caption{Function-code-docstring dataset statistics by language.}
\label{tab:func_docstring_stats}
\end{table}

\subsection{Class-Code-Docstring Dataset}

We extracted classes with associated docstrings using the same tree-sitter-based parsing pipeline. The initial scan identified over 300,000 classes across six languages (C is excluded as it lacks native class constructs). Table~\ref{tab:class_docstring_stats} presents the extraction statistics.

\begin{table}[htbp]
\centering
\begin{tabular}{@{}lrrr@{}}
\toprule
\textbf{Language} & \textbf{Total Classes} & \textbf{With Docstrings} & \textbf{Percentage} \\
\midrule
Matlab & 937 & 246 & 26.25\% \\
Fortran & 14,475 & 8,892 & 61.43\% \\
JavaScript & 4,413 & 1,453 & 32.93\% \\
Java & 62,668 & 38,288 & 61.10\% \\
Python & 186,546 & 85,556 & 45.86\% \\
C++ & 46,935 & 15,856 & 33.78\% \\
\midrule
\textbf{Total} & \textbf{315,974} & \textbf{150,291} & \textbf{47.56\%} \\
\bottomrule
\end{tabular}
\vspace{1ex}
\caption{Class-code-docstring dataset statistics by language.}
\label{tab:class_docstring_stats}
\end{table}

\subsection{Function-Code-Identifier Dataset}

For identifier-based retrieval, we applied additional filtering to ensure high-quality, semantically meaningful function names. The initial parsing extracted 3,403,419 functions. We applied the following filtering pipeline:

\begin{itemize}[noitemsep]
    \item \textbf{Length Filter}: Identifiers with fewer than 5 characters removed
    \item \textbf{Dictionary Filter}: snake\_case and camelCase identifier parts validated against an English dictionary
    \item \textbf{Frequency Filter}: Top 1\% most common identifiers removed to eliminate generic patterns
    \item \textbf{Stopword Filter}: Generic terms removed (e.g., \texttt{get}, \texttt{set}, \texttt{test}, \texttt{handler}, \texttt{init}, \texttt{main})
\end{itemize}

After filtering, function names in the code body are replaced with \texttt{<MASKED\_IDENTIFIER>} to prevent trivial keyword matching. The final dataset contains \textbf{761,136} high-quality samples.

\subsection{Class-Code-Identifier Dataset}

We applied the same filtering pipeline to class identifiers. The initial parsing extracted 502,518 classes across six languages. After applying length, dictionary, frequency, and stopword filters, class names in the code body are replaced with \texttt{<MASKED\_IDENTIFIER>}. The final dataset contains \textbf{172,365} high-quality samples.

\subsection{Rationale for Identifier Masking}

The identifier masking strategy serves several purposes in evaluating code retrieval systems:

\begin{itemize}[noitemsep]
    \item \textbf{Prevents trivial matching}: Without masking, retrieval systems could achieve high scores through simple keyword matching between the query (identifier) and code body
    \item \textbf{Forces semantic understanding}: Systems must understand code structure, variable usage patterns, and algorithmic behavior to match queries to code
    \item \textbf{Realistic evaluation}: Mirrors real-world scenarios where developers search for code functionality without knowing exact naming conventions
\end{itemize}

\section{Repository Search Benchmark Results}
\label{app:full_results}

Table~\ref{tab:repo_results_full} presents the complete repository search benchmark results across all evaluation cutoffs (@1, @5, and @10) for MRR, Recall, and NDCG metrics, reporting results for four retrieval approaches: BM25 (lexical baseline), all-MiniLM-L6-v2 (general-purpose sentence transformer), nasa-smd-ibm-st-v2 (domain-adapted embedding model), and indus-sde-st-v0.2 (Science Discovery Engine embedding model). Results are organized by text representation (rows) and scientific domain, with italic \textit{Hol.} rows indicating corpus-wide performance.

% --- SUBTABLE A: MRR ---
\begin{table*}[htbp]
\centering
\scriptsize
\begin{tabular}{@{}ll|ccc|ccc|ccc|ccc@{}}
\toprule
& & \multicolumn{3}{c|}{\textbf{BM25}} & \multicolumn{3}{c|}{\textbf{all-MiniLM-L6-v2}} & \multicolumn{3}{c|}{\textbf{nasa-smd-ibm-st-v2}} & \multicolumn{3}{c}{\textbf{indus-sde-st-v0.2}} \\
\textbf{View} & \textbf{Dom.} & @1 & @5 & @10 & @1 & @5 & @10 & @1 & @5 & @10 & @1 & @5 & @10 \\
\midrule
\multirow{4}{*}{\texttt{readme}}
& Earth & .18 & .22 & .23 & .29 & .34 & .36 & .33 & .42 & .43 & .27 & .37 & .39 \\
& Astro & .52 & .64 & .64 & .61 & .67 & .68 & .65 & .76 & .76 & .65 & .71 & .72 \\
& Plan. & .11 & .12 & .13 & .11 & .17 & .20 & .11 & .20 & .22 & .04 & .14 & .16 \\
& \textit{Hol.} & .22 & .27 & .28 & .31 & .37 & .39 & .34 & .44 & .45 & .30 & .39 & .41 \\
\midrule
\multirow{4}{*}{\texttt{cleaned}}
& Earth & .25 & .32 & .33 & .28 & .38 & .40 & .38 & .49 & .50 & .34 & .46 & .47 \\
& Astro & .52 & .65 & .66 & .65 & .74 & .75 & .61 & .69 & .69 & .65 & .75 & .76 \\
& Plan. & .11 & .14 & .15 & .19 & .22 & .25 & .11 & .14 & .16 & .11 & .23 & .26 \\
& \textit{Hol.} & .27 & .35 & .36 & .32 & .41 & .43 & .38 & .47 & .49 & .35 & .47 & .49 \\
\midrule
\multirow{4}{*}{\texttt{+topics}}
& Earth & .18 & .22 & .23 & .29 & .34 & .36 & .33 & .42 & .43 & .27 & .37 & .39 \\
& Astro & .52 & .64 & .64 & .61 & .67 & .68 & .65 & .76 & .76 & .65 & .71 & .72 \\
& Plan. & .11 & .12 & .13 & .11 & .17 & .20 & .11 & .20 & .22 & .04 & .14 & .16 \\
& \textit{Hol.} & .22 & .27 & .28 & .31 & .37 & .39 & .34 & .44 & .45 & .30 & .39 & .41 \\
\midrule
\multirow{4}{*}{\texttt{+context}}
& Earth & .16 & .21 & .22 & .28 & .34 & .35 & .34 & .42 & .43 & .27 & .37 & .38 \\
& Astro & .55 & .61 & .63 & .61 & .68 & .68 & .65 & .77 & .77 & .55 & .67 & .69 \\
& Plan. & .15 & .17 & .18 & .11 & .17 & .19 & .07 & .13 & .15 & .00 & .12 & .15 \\
& \textit{Hol.} & .21 & .26 & .27 & .31 & .37 & .38 & .35 & .43 & .44 & .28 & .38 & .40 \\
\midrule
\multirow{4}{*}{\texttt{comb.}$^\dagger$}
& Earth & .25 & .32 & .33 & .33 & .42 & .42 & .41 & .50 & .52 & .38 & .49 & .51 \\
& Astro & .65 & .75 & .75 & .71 & .82 & .82 & .84 & .87 & .87 & .71 & .82 & .83 \\
& Plan. & .11 & .15 & .16 & .19 & .24 & .26 & .15 & .19 & .22 & .11 & .23 & .24 \\
& \textit{Hol.} & .29 & .36 & .37 & .36 & .45 & .46 & .44 & .52 & .53 & .40 & .51 & .52 \\
\bottomrule
\end{tabular}
\vspace{1ex}
\caption{Full repository search benchmark results: MRR @1, @5, and @10 cutoffs. $^\dagger$\texttt{combined} concatenates \texttt{description}, \texttt{readme\_cleaned}, \texttt{topics}, and \texttt{additional\_context}.}
\label{tab:repo_results_full}
\end{table*}

% --- SUBTABLE B: RECALL ---
\begin{table*}[htbp]
\ContinuedFloat
\centering
\scriptsize
\begin{tabular}{@{}ll|ccc|ccc|ccc|ccc@{}}
\toprule
& & \multicolumn{3}{c|}{\textbf{BM25}} & \multicolumn{3}{c|}{\textbf{all-MiniLM-L6-v2}} & \multicolumn{3}{c|}{\textbf{nasa-smd-ibm-st-v2}} & \multicolumn{3}{c}{\textbf{indus-sde-st-v0.2}} \\
\textbf{View} & \textbf{Dom.} & @1 & @5 & @10 & @1 & @5 & @10 & @1 & @5 & @10 & @1 & @5 & @10 \\
\midrule
\multirow{4}{*}{\texttt{readme}}
& Earth & .18 & .28 & .37 & .29 & .44 & .55 & .33 & .56 & .62 & .27 & .52 & .66 \\
& Astro & .45 & .71 & .73 & .46 & .66 & .74 & .52 & .83 & .86 & .51 & .72 & .85 \\
& Plan. & .09 & .14 & .21 & .11 & .19 & .39 & .06 & .32 & .44 & .04 & .25 & .47 \\
& \textit{Hol.} & .21 & .32 & .40 & .29 & .44 & .56 & .32 & .57 & .63 & .28 & .51 & .66 \\
\midrule
\multirow{4}{*}{\texttt{cleaned}}
& Earth & .25 & .43 & .50 & .28 & .53 & .62 & .38 & .63 & .73 & .34 & .63 & .72 \\
& Astro & .43 & .64 & .74 & .47 & .77 & .84 & .46 & .71 & .75 & .49 & .78 & .87 \\
& Plan. & .09 & .20 & .24 & .16 & .25 & .44 & .06 & .19 & .33 & .09 & .36 & .58 \\
& \textit{Hol.} & .26 & .44 & .50 & .29 & .53 & .63 & .35 & .59 & .68 & .33 & .62 & .72 \\
\midrule
\multirow{4}{*}{\texttt{+topics}}
& Earth & .18 & .28 & .37 & .29 & .44 & .55 & .33 & .56 & .62 & .27 & .52 & .66 \\
& Astro & .45 & .71 & .73 & .46 & .66 & .74 & .52 & .83 & .86 & .51 & .72 & .85 \\
& Plan. & .09 & .14 & .21 & .11 & .19 & .39 & .06 & .32 & .44 & .04 & .25 & .47 \\
& \textit{Hol.} & .21 & .32 & .40 & .29 & .44 & .56 & .32 & .57 & .63 & .28 & .51 & .66 \\
\midrule
\multirow{4}{*}{\texttt{+context}}
& Earth & .16 & .29 & .35 & .28 & .44 & .55 & .34 & .55 & .60 & .27 & .53 & .63 \\
& Astro & .43 & .60 & .72 & .46 & .69 & .74 & .52 & .82 & .87 & .39 & .74 & .88 \\
& Plan. & .12 & .21 & .28 & .11 & .19 & .35 & .02 & .24 & .41 & .00 & .29 & .46 \\
& \textit{Hol.} & .19 & .32 & .40 & .29 & .44 & .56 & .32 & .55 & .62 & .25 & .53 & .65 \\
\midrule
\multirow{4}{*}{\texttt{comb.}$^\dagger$}
& Earth & .25 & .42 & .48 & .32 & .57 & .62 & .41 & .65 & .76 & .38 & .67 & .75 \\
& Astro & .51 & .71 & .76 & .53 & .88 & .89 & .64 & .86 & .93 & .53 & .82 & .93 \\
& Plan. & .09 & .24 & .31 & .16 & .29 & .43 & .09 & .26 & .43 & .09 & .42 & .52 \\
& \textit{Hol.} & .27 & .44 & .50 & .33 & .58 & .63 & .40 & .63 & .74 & .37 & .66 & .75 \\
\bottomrule
\end{tabular}
\vspace{1ex}
\caption{(continued) Recall @1, @5, and @10 cutoffs.}
\end{table*}

% --- SUBTABLE C: NDCG ---
\begin{table*}[htbp]
\ContinuedFloat
\centering
\scriptsize
\begin{tabular}{@{}ll|ccc|ccc|ccc|ccc@{}}
\toprule
& & \multicolumn{3}{c|}{\textbf{BM25}} & \multicolumn{3}{c|}{\textbf{all-MiniLM-L6-v2}} & \multicolumn{3}{c|}{\textbf{nasa-smd-ibm-st-v2}} & \multicolumn{3}{c}{\textbf{indus-sde-st-v0.2}} \\
\textbf{View} & \textbf{Dom.} & @1 & @5 & @10 & @1 & @5 & @10 & @1 & @5 & @10 & @1 & @5 & @10 \\
\midrule
\multirow{4}{*}{\texttt{readme}}
& Earth & .18 & .24 & .26 & .29 & .37 & .41 & .33 & .45 & .47 & .27 & .40 & .45 \\
& Astro & .52 & .62 & .63 & .61 & .62 & .65 & .65 & .76 & .76 & .65 & .69 & .72 \\
& Plan. & .11 & .12 & .14 & .11 & .17 & .23 & .11 & .22 & .25 & .04 & .16 & .23 \\
& \textit{Hol.} & .22 & .28 & .30 & .31 & .38 & .42 & .34 & .47 & .49 & .30 & .41 & .46 \\
\midrule
\multirow{4}{*}{\texttt{cleaned}}
& Earth & .25 & .35 & .37 & .28 & .42 & .45 & .38 & .53 & .56 & .34 & .50 & .53 \\
& Astro & .52 & .59 & .62 & .65 & .71 & .73 & .61 & .65 & .67 & .65 & .73 & .76 \\
& Plan. & .11 & .14 & .15 & .19 & .22 & .29 & .11 & .13 & .18 & .11 & .25 & .32 \\
& \textit{Hol.} & .27 & .36 & .38 & .32 & .44 & .47 & .38 & .49 & .53 & .35 & .50 & .54 \\
\midrule
\multirow{4}{*}{\texttt{+topics}}
& Earth & .18 & .24 & .26 & .29 & .37 & .41 & .33 & .45 & .47 & .27 & .40 & .45 \\
& Astro & .52 & .62 & .63 & .61 & .62 & .65 & .65 & .76 & .76 & .65 & .69 & .72 \\
& Plan. & .11 & .12 & .14 & .11 & .17 & .23 & .11 & .22 & .25 & .04 & .16 & .23 \\
& \textit{Hol.} & .22 & .28 & .30 & .31 & .38 & .42 & .34 & .47 & .49 & .30 & .41 & .46 \\
\midrule
\multirow{4}{*}{\texttt{+context}}
& Earth & .16 & .23 & .25 & .28 & .36 & .40 & .34 & .45 & .47 & .27 & .41 & .44 \\
& Astro & .55 & .58 & .61 & .61 & .64 & .65 & .65 & .76 & .77 & .55 & .65 & .70 \\
& Plan. & .15 & .17 & .19 & .11 & .17 & .22 & .07 & .14 & .20 & .00 & .16 & .22 \\
& \textit{Hol.} & .21 & .27 & .30 & .31 & .38 & .41 & .35 & .46 & .48 & .28 & .41 & .45 \\
\midrule
\multirow{4}{*}{\texttt{comb.}$^\dagger$}
& Earth & .25 & .34 & .36 & .33 & .45 & .47 & .41 & .54 & .57 & .38 & .54 & .56 \\
& Astro & .65 & .68 & .69 & .71 & .80 & .80 & .84 & .83 & .85 & .71 & .77 & .81 \\
& Plan. & .11 & .16 & .18 & .19 & .24 & .29 & .15 & .19 & .25 & .11 & .26 & .30 \\
& \textit{Hol.} & .29 & .37 & .39 & .36 & .48 & .49 & .44 & .54 & .57 & .40 & .54 & .57 \\
\bottomrule
\end{tabular}
\vspace{1ex}
\caption{(continued) NDCG @1, @5, and @10 cutoffs.}
\end{table*}

\clearpage
\section{Dataset and Code Availability}
\label{app:access}

All three datasets are publicly available on HuggingFace:

\begin{itemize}
    \item \textbf{Repository Corpus (5,264 repos)}: \\ \url{https://huggingface.co/datasets/nasa-impact/nasa-science-github-repos}

    \item \textbf{Repository Search Benchmark (219 queries)}: \\ \url{https://huggingface.co/datasets/nasa-impact/nasa-science-repos-sme-benchmark}

    \item \textbf{Code Snippet Retrieval Benchmark}: \\ \url{https://huggingface.co/datasets/nasa-impact/nasa-science-code-benchmark-v0.1}
\end{itemize}

The accompanying code is available on GitHub:
\begin{itemize}
    \item \textbf{Benchmarking Code}: Full evaluation scripts for replicability \\ \url{https://github.com/NASA-IMPACT/nasa-science-repo-benchmark}

    \item \textbf{Data Collection Pipeline}: Corpus aggregation and processing \\ \url{https://github.com/NASA-IMPACT/github-code-discovery}
\end{itemize}

\begin{table*}[htbp]
\centering
\scriptsize

% --- SUBTABLE A: MRR ---
\begin{subtable}{\textwidth}
    \centering
    \resizebox{\textwidth}{!}{
    \begin{tabular}{@{}ll|ccc|ccc|ccc|ccc|ccc@{}}
    \toprule
    & & \multicolumn{3}{c|}{\textbf{BM25}} & \multicolumn{3}{c|}{\textbf{indus-sde-st-v0.2}} & \multicolumn{3}{c|}{\textbf{nasa-smd-ibm-st-v2}} & \multicolumn{3}{c|}{\textbf{Qwen3-Embedding-0.6B}} & \multicolumn{3}{c}{\textbf{SFR-Embedding-Code-400M\_R}} \\
    \textbf{Group} & \textbf{Category} & @1 & @5 & @10 & @1 & @5 & @10 & @1 & @5 & @10 & @1 & @5 & @10 & @1 & @5 & @10 \\
    \midrule
    \multirow{5}{*}{\textbf{Division}}
    & Astro & .14 & .18 & .18 & .19 & .25 & .26 & .26 & .33 & .33 & .43 & .50 & .50 & .30 & .35 & .36 \\
    & Bio. \& Phys. & .09 & .12 & .12 & .17 & .24 & .25 & .21 & .28 & .29 & .50 & .57 & .57 & .24 & .30 & .30 \\
    & Earth & .17 & .21 & .22 & .23 & .30 & .31 & .32 & .39 & .39 & .51 & .58 & .58 & .37 & .43 & .43 \\
    & Helio & .22 & .26 & .27 & .26 & .32 & .33 & .31 & .38 & .39 & .47 & .53 & .53 & .37 & .43 & .44 \\
    & Planetary & .15 & .19 & .20 & .19 & .25 & .26 & .26 & .33 & .34 & .51 & .59 & .59 & .34 & .41 & .42 \\
    \midrule
    \multirow{4}{*}{\shortstack[l]{\textbf{Query}\\\textbf{Type}}}
    & Class Doc & .24 & .29 & .30 & .29 & .36 & .37 & .41 & .50 & .51 & .66 & .74 & .74 & .53 & .62 & .63 \\
    & Class ID & .01 & .01 & .01 & .09 & .13 & .14 & .11 & .16 & .16 & .18 & .24 & .25 & .09 & .13 & .14 \\
    & Func Doc & .23 & .29 & .30 & .27 & .35 & .36 & .37 & .46 & .47 & .68 & .75 & .76 & .47 & .55 & .56 \\
    & Func ID & .00 & .01 & .01 & .07 & .11 & .12 & .08 & .12 & .13 & .12 & .17 & .18 & .04 & .05 & .06 \\
    \midrule
    \multirow{7}{*}{\shortstack[l]{\textbf{Progr.}\\\textbf{Lang.}}}
    & C++ & .14 & .17 & .18 & .21 & .28 & .29 & .26 & .34 & .35 & .59 & .66 & .66 & .32 & .39 & .40 \\
    & C & .16 & .20 & .21 & .18 & .24 & .25 & .26 & .33 & .34 & .55 & .63 & .63 & .32 & .38 & .39 \\
    & Fortran & .14 & .17 & .18 & .21 & .27 & .28 & .23 & .29 & .30 & .42 & .48 & .48 & .28 & .33 & .34 \\
    & Java & .03 & .04 & .05 & .14 & .19 & .20 & .16 & .21 & .22 & .25 & .32 & .33 & .13 & .18 & .18 \\
    & Javascript & .23 & .27 & .27 & .22 & .29 & .30 & .29 & .36 & .36 & .56 & .62 & .62 & .39 & .44 & .45 \\
    & Matlab & .03 & .04 & .04 & .12 & .15 & .15 & .17 & .21 & .21 & .22 & .25 & .26 & .18 & .21 & .21 \\
    & Python & .16 & .21 & .21 & .22 & .28 & .28 & .30 & .37 & .37 & .46 & .52 & .53 & .35 & .41 & .41 \\
    \midrule
    \textbf{Holistic} & Overall & .15 & .18 & .19 & .20 & .26 & .27 & .27 & .34 & .34 & .47 & .54 & .54 & .31 & .37 & .38 \\
    \bottomrule
    \end{tabular}
    }
    \caption{MRR @1, @5, and @10 cutoffs.}
\end{subtable}

\vspace{2ex}

% --- SUBTABLE B: RECALL ---
\begin{subtable}{\textwidth}
    \centering
    \resizebox{\textwidth}{!}{
    \begin{tabular}{@{}ll|ccc|ccc|ccc|ccc|ccc@{}}
    \toprule
    & & \multicolumn{3}{c|}{\textbf{BM25}} & \multicolumn{3}{c|}{\textbf{indus-sde-st-v0.2}} & \multicolumn{3}{c|}{\textbf{nasa-smd-ibm-st-v2}} & \multicolumn{3}{c|}{\textbf{Qwen3-Embedding-0.6B}} & \multicolumn{3}{c}{\textbf{SFR-Embedding-Code-400M\_R}} \\
    \textbf{Group} & \textbf{Category} & @1 & @5 & @10 & @1 & @5 & @10 & @1 & @5 & @10 & @1 & @5 & @10 & @1 & @5 & @10 \\
    \midrule
    \multirow{5}{*}{\textbf{Division}}
    & Astro & .14 & .23 & .26 & .19 & .35 & .41 & .26 & .43 & .49 & .43 & .60 & .64 & .30 & .44 & .48 \\
    & Bio. \& Phys. & .09 & .16 & .19 & .17 & .34 & .41 & .21 & .39 & .47 & .50 & .67 & .71 & .24 & .39 & .44 \\
    & Earth & .17 & .27 & .31 & .23 & .41 & .47 & .32 & .49 & .55 & .51 & .67 & .72 & .37 & .52 & .57 \\
    & Helio & .22 & .33 & .36 & .26 & .43 & .49 & .31 & .49 & .54 & .47 & .60 & .63 & .37 & .53 & .56 \\
    & Planetary & .15 & .26 & .30 & .19 & .35 & .41 & .26 & .45 & .52 & .51 & .70 & .74 & .34 & .53 & .58 \\
    \midrule
    \multirow{4}{*}{\shortstack[l]{\textbf{Query}\\\textbf{Type}}}
    & Class Doc & .24 & .38 & .44 & .29 & .49 & .56 & .41 & .63 & .70 & .66 & .85 & .88 & .53 & .76 & .80 \\
    & Class ID & .01 & .02 & .02 & .09 & .20 & .25 & .11 & .23 & .29 & .18 & .35 & .41 & .09 & .20 & .25 \\
    & Func Doc & .23 & .38 & .43 & .27 & .47 & .54 & .37 & .59 & .66 & .68 & .86 & .89 & .47 & .67 & .72 \\
    & Func ID & .00 & .01 & .01 & .07 & .18 & .22 & .08 & .19 & .24 & .12 & .26 & .32 & .04 & .09 & .12 \\
    \midrule
    \multirow{7}{*}{\shortstack[l]{\textbf{Progr.}\\\textbf{Lang.}}}
    & C++ & .14 & .24 & .28 & .21 & .40 & .47 & .26 & .46 & .54 & .59 & .76 & .80 & .32 & .49 & .55 \\
    & C & .16 & .27 & .30 & .18 & .34 & .41 & .26 & .45 & .52 & .55 & .74 & .79 & .32 & .48 & .54 \\
    & Fortran & .14 & .23 & .27 & .21 & .39 & .45 & .23 & .40 & .47 & .42 & .57 & .60 & .28 & .41 & .45 \\
    & Java & .03 & .06 & .07 & .14 & .28 & .35 & .16 & .31 & .37 & .25 & .43 & .49 & .13 & .25 & .30 \\
    & Javascript & .23 & .33 & .37 & .22 & .39 & .45 & .29 & .46 & .51 & .56 & .70 & .74 & .39 & .51 & .54 \\
    & Matlab & .03 & .04 & .04 & .12 & .20 & .22 & .17 & .28 & .32 & .22 & .31 & .35 & .18 & .25 & .28 \\
    & Python & .16 & .27 & .30 & .22 & .38 & .43 & .30 & .47 & .53 & .46 & .62 & .66 & .35 & .50 & .54 \\
    \midrule
    \textbf{Holistic} & Overall & .15 & .24 & .27 & .20 & .36 & .43 & .27 & .45 & .51 & .47 & .64 & .68 & .31 & .47 & .51 \\
    \bottomrule
    \end{tabular}
    }
    \caption{Recall @1, @5, and @10 cutoffs.}
\end{subtable}

\vspace{2ex}

% --- SUBTABLE C: NDCG ---
\begin{subtable}{\textwidth}
    \centering
    \resizebox{\textwidth}{!}{
    \begin{tabular}{@{}ll|ccc|ccc|ccc|ccc|ccc@{}}
    \toprule
    & & \multicolumn{3}{c|}{\textbf{BM25}} & \multicolumn{3}{c|}{\textbf{indus-sde-st-v0.2}} & \multicolumn{3}{c|}{\textbf{nasa-smd-ibm-st-v2}} & \multicolumn{3}{c|}{\textbf{Qwen3-Embedding-0.6B}} & \multicolumn{3}{c}{\textbf{SFR-Embedding-Code-400M\_R}} \\
    \textbf{Group} & \textbf{Category} & @1 & @5 & @10 & @1 & @5 & @10 & @1 & @5 & @10 & @1 & @5 & @10 & @1 & @5 & @10 \\
    \midrule
    \multirow{5}{*}{\textbf{Division}}
    & Astro & .14 & .19 & .20 & .19 & .27 & .29 & .26 & .35 & .37 & .43 & .52 & .54 & .30 & .37 & .39 \\
    & Bio. \& Phys. & .09 & .13 & .14 & .17 & .26 & .29 & .21 & .30 & .33 & .50 & .59 & .61 & .24 & .32 & .34 \\
    & Earth & .17 & .23 & .24 & .23 & .33 & .35 & .32 & .41 & .43 & .51 & .60 & .61 & .37 & .45 & .47 \\
    & Helio & .22 & .28 & .29 & .26 & .35 & .37 & .31 & .41 & .43 & .47 & .54 & .55 & .37 & .46 & .47 \\
    & Planetary & .15 & .21 & .22 & .19 & .27 & .30 & .26 & .36 & .39 & .51 & .61 & .63 & .34 & .44 & .46 \\
    \midrule
    \multirow{4}{*}{\shortstack[l]{\textbf{Query}\\\textbf{Type}}}
    & Class Doc & .24 & .31 & .34 & .29 & .40 & .42 & .41 & .53 & .55 & .66 & .77 & .78 & .53 & .65 & .67 \\
    & Class ID & .01 & .01 & .01 & .09 & .15 & .16 & .11 & .18 & .19 & .18 & .27 & .29 & .09 & .15 & .16 \\
    & Func Doc & .23 & .31 & .33 & .27 & .38 & .40 & .37 & .49 & .51 & .68 & .78 & .79 & .47 & .58 & .60 \\
    & Func ID & .00 & .01 & .01 & .07 & .13 & .14 & .08 & .14 & .15 & .12 & .19 & .21 & .04 & .06 & .07 \\
    \midrule
    \multirow{7}{*}{\shortstack[l]{\textbf{Progr.}\\\textbf{Lang.}}}
    & C++ & .14 & .19 & .20 & .21 & .31 & .34 & .26 & .37 & .39 & .59 & .68 & .69 & .32 & .41 & .43 \\
    & C & .16 & .22 & .23 & .18 & .26 & .28 & .26 & .36 & .38 & .55 & .65 & .67 & .32 & .41 & .42 \\
    & Fortran & .14 & .18 & .20 & .21 & .30 & .32 & .23 & .32 & .34 & .42 & .50 & .51 & .28 & .35 & .36 \\
    & Java & .03 & .05 & .05 & .14 & .21 & .23 & .16 & .24 & .26 & .25 & .35 & .37 & .13 & .20 & .21 \\
    & Javascript & .23 & .28 & .30 & .22 & .31 & .33 & .29 & .38 & .40 & .56 & .64 & .65 & .39 & .46 & .47 \\
    & Matlab & .03 & .04 & .04 & .12 & .16 & .17 & .17 & .23 & .24 & .22 & .27 & .28 & .18 & .22 & .23 \\
    & Python & .16 & .22 & .23 & .22 & .30 & .32 & .30 & .39 & .41 & .46 & .55 & .56 & .35 & .43 & .44 \\
    \midrule
    \textbf{Holistic} & Overall & .15 & .20 & .21 & .20 & .29 & .31 & .27 & .36 & .38 & .47 & .56 & .58 & .31 & .40 & .41 \\
    \bottomrule
    \end{tabular}
    }
    \caption{NDCG @1, @5, and @10 cutoffs.}
\end{subtable}

\centering
\vspace{1ex}
\caption{Code search evaluation results across Divisions, Query Types, and Programming Languages.}
\label{tab:consolidated_results}
\end{table*}

\onecolumn
\section{Full Division Classification Prompt}
\label{app:division_prompt}

The following is the complete system prompt used for LLM-based division classification. The prompt is constructed from three components: (1) the instruction template, (2) evaluation criteria for each NASA SMD division, and (3) few-shot classification examples.

\textbf{Instruction Template.}

\begin{lstlisting}[basicstyle=\ttfamily\small, breaklines=true, columns=fullflexible]
instructions = f"""
You are an expert assistant specialized in classifying technical README files according to NASA's research divisions. Your task is to analyze the provided README content and determine which NASA research area it primarily belongs to. You must use the EVALUATION CRITERIA provided below. If there are signals or indirect references that suggest alignment with a NASA research area, classify accordingly and provide reasoning.

EVALUATION CRITERIA: {evaluation_criteria}

Examples: {few_shot_examples}
"""
\end{lstlisting}

\textbf{Evaluation Criteria} (\texttt{evaluation\_criteria}).

\begin{lstlisting}[basicstyle=\ttfamily\small, breaklines=true, columns=fullflexible]
Earth Science Division
Overview
NASA's Earth Science Division develops and operates satellite, airborne, and ground-based programs to observe and analyze Earth's atmosphere, oceans, land, ice sheets, and ecosystems in order to understand climate dynamics, natural hazards, and environmental change.
Study Areas & Examples
Agriculture & Water Cycle Monitoring
Soil moisture and precipitation studies using SMAP and GRACE missions.
Carbon Cycle & Atmospheric Composition
Tracking greenhouse gases with the Orbiting Carbon Observatory-2 (OCO-2).
Sea-Level & Cryosphere Dynamics
Measuring ocean height and ice-sheet elevations with Sentinel-6/Jason CS and ICESat-2.
Land Cover & Ecosystem Change
Assessing vegetation and land-use via MODIS instruments on Terra and Aqua.
Disaster Preparedness & Response
Supporting flood, wildfire, and hurricane monitoring through the GOES weather satellites.


Planetary Science Division
Overview
NASA's Planetary Science Division explores planets, moons, asteroids, and comets throughout the solar system via robotic spacecraft, sample returns, and telescopic observations to unravel its formation history and search for signs of past or present life.
Study Areas & Examples
Inner Solar System Exploration
MESSENGER at Mercury, Magellan at Venus, and Lunar Reconnaissance Orbiter at the Moon.
Mars Habitability & Geology
Rovers Curiosity and Perseverance, and the InSight lander studying Martian surface and interior.
Outer Planets & Ocean Worlds
Juno at Jupiter, Cassini at Saturn, and the forthcoming Europa Clipper mission.
Small Bodies & Sample Return
OSIRIS-REx (asteroid Bennu), Hayabusa2 (asteroid Ryugu), Lucy (Trojan asteroids), and New Horizons (Pluto).
Planetary Defense
Detecting and tracking near-Earth objects with NEOWISE and coordinating response via the Planetary Defense Coordination Office.


Astrophysics Division
Overview
NASA's Astrophysics Division seeks to understand the universe's origin, structure, evolution, and potential for life by deploying space observatories and supporting theoretical research to address fundamental cosmic questions.
Study Areas & Examples
Cosmic Origins
Mapping early galaxies and star formation with Hubble's Cosmic Origins Spectrograph.
Physics of the Cosmos
Investigating dark matter, dark energy, and black holes with the Chandra X-ray Observatory.
Exoplanet Exploration
Discovering and characterizing exoplanets using Kepler and TESS missions.
Flagship Observatories
Operating large telescopes--Hubble and James Webb--to observe deep-space phenomena.


Heliophysics Division
Overview
NASA's Heliophysics Division studies the Sun, solar wind, and heliosphere to understand space weather, magnetic reconnection, and their impacts on planetary environments and technology.
Study Areas & Examples
Solar Dynamics
Investigating the solar corona and wind acceleration with Parker Solar Probe and Solar Dynamics Observatory.
Space Weather & Magnetospheres
Monitoring geomagnetic storms and radiation belts using Van Allen Probes and the Magnetospheric Multiscale Mission (MMS).
Heliosphere & Interstellar Boundary
Mapping the heliosphere's edge with Voyager spacecraft and IBEX.
Heliophysics System Observatory
Coordinating a fleet of missions to study solar-terrestrial interactions across the solar system.


Biological and Physical Sciences Division
Overview
NASA's Biological and Physical Sciences Division leverages microgravity and space radiation to conduct fundamental research in life sciences and physical sciences, supporting long-duration space exploration and improving life on Earth.
Study Areas & Examples
Space Biology
Studying molecular, cellular, plant, animal, and human biology aboard the ISS to understand microgravity effects.
Physical Sciences
Investigating biophysics, combustion, fluid dynamics, materials science, and fundamental physics in space.
Technology & Applications
Developing quantum sensors, atomic clocks, and tissue-chip systems for both spaceflight and Earth applications.
Data & Open Science
Sharing results via open platforms like GeneLab and the Physical Sciences Informatics System (PSI).
\end{lstlisting}

\textbf{Few-Shot Examples} (\texttt{few\_shot\_examples}).

\begin{lstlisting}[basicstyle=\ttfamily\small, breaklines=true, columns=fullflexible]
--- Example 1 ---
Input README Snippet:
# Object-Based Image Analysis (OBIA) and Machine Learning (ML) Applied to High Spatial Accuracy Forest Mapping in Parana, Southern Brazil This repository organizes the Mapped codes for Machine Learning in GEE Requisites: Google Earth Engine (GEE)
Expected Classification:
{"area": "Earth Science Division"}

--- Example 2 ---
Input README Snippet:
#Data and code generated for DryFlux
#Final Submission to Nature Communications Earth & Environment on Oct2021
Ecohydrological water-carbon coupling improves dryland carbon flux
prediction of average uptake, interannual variability, and drought
Authors: Barnes, Mallory L.; Farella, Martha M.; Scott, Russell L.;
Moore, David J.P.; Ponce-Campos. Guillermo E.; Biederman, Joel A.;
MacBean, Natasha; Litvak, Marcy E.; and Breshears, David D.
Year: 2021
Title: Data and code for DryFlux
Corresponding Author for Code: Martha Farella, Indiana University,
O'Neill School of Public and Environmental Affairs, [email redacted]
License: MIT
DOI:
---------------------------------------------
## Summary
Contains code and data used for DryFlux and the analysis presented in
Nature Communications Earth & Environment manuscript, Ecohydrological
water-carbon coupling improves dryland carbon flux prediction of average
uptake, interannual variability, and drought
---------------------------------------------
## Files and Folders
code: code used in the analysis presented in the manuscript.
MATMAPelev.R: get metadata for the Flux tower sites.
1.download SRTM elevation tiles and extract values for site locations
2. download WorldClim MAT/MAP tiles and extract data for site locations;
3. combine 1 and 2 into a single dataframe;
4. determine the OzFlux sites that fall within the 'dryland regions'
defined by United Nations Environment World Monitoring Centre;
5. get boundaries for US, mexico, and Australia territories
SPEI: code used to compute SPEI and extract meterological variables for
flux tower locations. CRU data downloaded from:
https://crudata.uea.ac.uk/cru/data/hrg/cru_ts_4.04/cruts.2004151855.v4.04/
meterological variables downloaded included: precipitation (pre),
tmin (tmn), tmax (tmx), vapor pressure (vap), potential
evapotranspiration (pet), and Tavg (tmp) for time periods: 1991-2000;
2001-2010; 2011-2019;
citation: Harris et al. (2020) doi:10.1038/s41597-020-0453-3
combineNCD.R: combine the seperate cru .nc files into a single file
that contains oberservations from 1991 - 2019 for each variable needed
to calculate spei (pet and precip)
computeSPEI.R: This script computes the global SPEI dataset at different
time scales. One netCDF file covering the whole globe and time period is
generated for each time scale, e.g. spei01.nc for a time scale of
1 month, etc. Output files are stored on DryFlux/data/outputNcdf
functions.R: dependencies for 'computeSPEI.R'
SPEIextract: code used to extract cru data values for flux tower
locations, pre-process for downstream analysis, and create rasters for
upscaling; lines 13:90 are for SPEI extraction; lines 92:144 are for
the other meterological variables; lines 146:177 are for last month
precip and Tavg; lines 179:340 are for OzFlux sites; lines 345:438
raster layer creation
MODISprocessing.R: used to convert hdf files to a single .csv file for
each year where ndvi and evi values are extracted for flux tower
locations and data formatted (single row/observation for each site/date)
for downstream analysis also create .tif EVI and NDVI raster layers;
.hdf data products downloaded from:
https://e4ftl01.cr.usgs.gov/MOLT/MOD13C1.006/
16th day global 0.5deg CMG MOD13C1; Date downloaded is the date closest
to the 15th/16th of each month;
citation: Didan, K. (2015). MOD13C1 MODIS/Terra Vegetation Indices
16-Day L3 Global 0.05Deg CMG V006 [Data set]. NASA EOSDIS Land
Processes DAAC. Accessed 2020-10-13 from
https://doi.org/10.5067/MODIS/MOD13C1.006
daylength.R: determine daylength of sites based on dates of CRU dates;
create daylength rasters; daylength calculated using the 'daylength'
function in the 'geosphere' R package; daylength calculated on the 15th
or 16th of each month in accordance with CRU dates for each flux tower
location.; raster stack created for the whole world with daylengths for
each of the 348 CRU dates
CleanFlux.R: get fluxtower GPP values for mid-month time frames from
the daily Fluxtower MatLab files
MOD17A2Hdl.txt: code used on google EarthEngine to download individual
.csv files of MODIS 8-day global 500m GPP data products (MOD17A2H) for
each flux tower site.
MODIS_GPP.R: get MODIS GPP values in format needed for downstream
analysis. MODIS 8-day global 500m GPP data products (MOD17A2H)
downloaded with google EarthEngine using code on 'MOD17A2Hdl.txt'.
Fluxcom_extract.R: extract daily Fluxcom predictions for tower locs,
then calc mid-month GPP vals; daily Fluxcom GPP predictions requested
from Fluxcom data administrator
OxFlux.R: get OzFlux GPP values formatted for downstream analysis
data_prepro.R: combine all response and explanatory variables together
into a single dataset for model building, testing, and analysis
RF.R: Build the Random Forest Machine Learning Models, save model
outputs and evaluate performance
applyRF_loop.R: Apply the random forest model to raster layers to
predict global GPP across all months
RFevaluation.R: Evaluate model performance. Create a dataframe with
observed and predicted GPP values from DryFlux, MODIS, and Fluxcom.
Create Fig1a-c, Fig2a-b, Table S2 and S5, Figs. S2, S3, S5, and S6
RFseasonal.R: Evaluate seasonal trends in RF predictions at SW USA and
OzFluxsites. Create Fig1d-f, Figs. S4 and S7
RFspatial_maps.R: calcualte mid-monthly composites of Fluxcom GPP
predictiosn, calculate annual GPP estimates for Fluxcom and DryFlux,
create annual difference maps (Fig2c,d; Fig3c,d), calcualte DryFlux GPP
z-scores globally and create maps (Figs3a,b)
data: data used in analysis presented in the manuscript
contents of this folder include:
site_locs.csv : site meta data for Southwestern NA sites including:
site ID, site name, lat/long, vegetation classification (determined from
MODIS IGBP land cover classification at flux tower location), MAT (from
FluxNet), MAP (from FluxNet), and elevation (from FluxNet)
global_locs.csv: site meta data from FluxNET for the global dryland sites
site_meta.csv: WorldClim MAP/MAT, and STRM elevation meta-data for SW
USA sites (output from 'MATMAPelev.R')
global_meta.csv: WorldClim MAP/MAT, and STRM elevation meta-data for
the global dryland sites (output from 'MATMAPelev.R')
dates.csv: dates of CRU data
RFdata.csv: output from 'data_prepro.R' all of the data needed for
Random Forest analysis on the SW USA sites.
global_RFdata.csv: output from 'data_prepro.R' all of the data needed
for Random Forest analysis on the Global dryland sites
Flux_raw: MatLab files for each fluxtower site containing daily NEP,
GEP, Reco, ET, precip, Tair, VPD, Rnet, and Rsolar
Expected Classification:
{"area": "Earth Science Division"}

--- Example 3 ---
Input README Snippet:
# GRIZZLY Details of the code can be found here https://arxiv.org/abs/1710.09397. This code has two part. First to generate the 1D profiles around different sources. Second is to use those around the dark matter halos in the simulation box and use the density and velocity fields to generate the brightness temperature maps. Author: Raghunath Ghara Date: 12 Jan 2019
Expected Classification:
{"area": "Astrophysics Division"}

--- Example 4 ---
Input README Snippet:
# rfpipe
A fast radio interferometric transient search library. Extends on
rtpipe. This library supports real-time analysis for the realfast
project and offline analysis of VLA data on a single workstation.
Integration with the real-time VLA and cluster processing is provided
by realfast.
Planned future development include:
- Supporting other search algorithms.
- Supporting other interferometers by adding data and metadata reading
functions.
## Installation
rfpipe requires the anaconda installer on Linux and OSX. The most
reliable installation is for Python3.6 and adding conda-forge:
conda config --add channels conda-forge
conda create -n realfast python=3.6 numpy scipy cython matplotlib
numba pyfftw bokeh
source activate realfast
pip install --extra-index-url
https://casa-pip.nrao.edu:443/repository/pypi-group/simple casatools
pip install -e
git+git://github.com/realfastvla/rfpipe#egg=rfpipe
## Dependencies
- numpy/scipy/matplotlib
- casa6 python libraries (for quanta and measures; available on
Python 3.6 via pip)
- numba (for multi-core and gpu acceleration)
- rtpipe (for flagging; will be removed soon)
- astropy
- sdmpy
- pyfftw
- pyyaml
- attrs
- rfgpu (optional; for GPU FFTs)
- vys/vysmaw and vysmaw_reader (optional; to read vys data from VLA
correlator)
## Citation
If you use rfpipe, please support open software by citing the record
on the Astrophysics Source Code Library at http://ascl.net/1710.002.
Expected Classification:
{"area": "Astrophysics Division"}
--- End of Examples ---
\end{lstlisting}
\end{document}